\documentclass[fleqn,usenatbib]{mnras}

\usepackage{mathptmx}

\usepackage{hyperref}
\hypersetup{colorlinks = false}

\usepackage[T1]{fontenc}
\usepackage{xspace}
\usepackage[english]{babel}
\usepackage{array}
\usepackage{tabularx}
\usepackage{booktabs}
\usepackage{hhline}
\usepackage{soul}
\usepackage[usenames]{color}
\usepackage{algpseudocode}
\usepackage{siunitx}
\usepackage{graphicx}
\usepackage{graphbox}
\usepackage{amsmath}
\usepackage{amssymb}
\usepackage{tikz}
\usepackage{physics}
\usepackage{cleveref}
\usepackage{acronym}
\usepackage{bm}
\usetikzlibrary{arrows,decorations.pathreplacing,positioning,calc}
\usepackage{ifthen}
\usetikzlibrary{shapes,backgrounds}
\crefname{equation}{equation}{equations}
\crefname{figure}{Fig.}{Figs.}

\Crefname{equation}{Equation}{Equations}
\Crefname{figure}{Figure}{Figures}

\crefname{appendix}{Appendix}{Appendices}
\Crefname{appendix}{Appendix}{Appendices}

\crefname{table}{Table}{Tables}
\crefrangelabelformat{equation}{(#3#1#4)--(#5#2#6)}

\def \apj {{\it ApJ}}
\def \apjl {{\it ApJ Let.}}
\def \apjs {{\it ApJ Sup.}}

\def \aa {{\it A\&A}}
\def \aap {{\it A\&A}}

\def \mnras {{\it MNRAS}}

\definecolor{grey}{rgb}{0.75,0.75,0.75}
\definecolor{Orange}{rgb}{1.0,0.5,0.15}
\definecolor{brown}{rgb}{0.7,0.25,0.0}
\definecolor{pink}{rgb}{1.0,0.5,0.5}
\definecolor{darkerred}{rgb}{0.8,0,0}
\definecolor{darkerblue}{rgb}{0,0,0.8}
\definecolor{Blue}{rgb}{0,0.08,0.65}
\definecolor{Red}{rgb}{0.65,0.08,0.05}
\definecolor{Green}{rgb}{0.15,0.45,0.25}

\newcommand{\Msun}{\ensuremath{M_\odot}}

\newcommand{\ramses}{{\sc ramses}\@}

\newcommand{\Rvir}{\ensuremath{R_\mathrm{vir}}}
\renewcommand{\vec}[1]{\bm{#1}}

\newcommand{\ie}{\emph{i.e.}\xspace}
\newcommand{\eg}{e.g.\xspace}

\newcommand{\rr}{{\vec{r}}}
\newcommand{\vv}{{\vec{v}}}
\graphicspath{{figures/}}
\newcommand{\adaptahop}{\textsc{adaptahop}}

\acrodef{AM}[AM]{angular momentum}
\newcommand{\AM}{\ac{AM}\xspace}
\acrodef{sAM}[sAM]{specific angular momentum}
\newcommand{\sAM}{\ac{sAM}\xspace}

\begin{document}

\author[C.~Cadiou, Y.~Dubois \& C.~Pichon]{Corentin~Cadiou$^{1,2}$ \thanks{c.cadiou@ucl.ac.uk}, Yohan~Dubois$^{2}$, and Christophe~Pichon$^{2,3}$\\
$^{1}$ Department of Physics and Astronomy, University College London, Gower Street, London WC1E 6BT, United-Kingdom \\
$^2$ CNRS and Sorbonne Université, UMR 7095, Institut d'Astrophysique de Paris, 98 bis Boulevard Arago, F-75014 Paris, France\\
$^3$ Université Paris-Saclay, CNRS, CEA, Institut de physique théorique, 91191, Gif-sur-Yvette, France.
}

\title[Angular momentum dynamics dominated by gravity]{
Gravitational torques dominate the dynamics %
of  accreted gas at $z>2$
}

\maketitle
{
 }
 \begin{abstract}
    Galaxies form from the accretion of cosmological infall of gas.
    In the high redshift Universe, most of this gas infall is expected to be dominated by cold filamentary flows which connect deep down inside  halos, and, hence, to the vicinity of  galaxies.
    Such cold flows are  important since they dominate the mass and angular momentum acquisition that can make up rotationally-supported disks at high-redshifts.
     We study the angular momentum acquisition of gas into galaxies, and in particular, the torques acting on the accretion flows, using
        hydrodynamical cosmological simulations of high-resolution zoomed-in halos of a few $10^{11}\,\rm M_\odot$  at $z=2$.
    Torques can be separated into those of gravitational origin, and   hydrodynamical ones driven by pressure gradients.
    We find that coherent gravitational torques dominate over pressure torques in the cold phase, and are hence responsible for the spin-down and realignment of this gas.
    Pressure torques display small-scale fluctuations of significant amplitude, but with very little coherence on the relevant galaxy or halo-scale that would otherwise allow them to effectively re-orientate the gas flows.
    Dark matter torques dominate gravitational torques outside the galaxy, while within the galaxy, the baryonic component dominates. The circum-galactic medium emerges as the transition region for angular momentum re-orientation of the cold component towards the central galaxy's mid-plane.
\end{abstract}

\begin{figure*}
\begin{center}
    \resizebox{0.9\textwidth}{!}{
        \includegraphics[height=2cm]{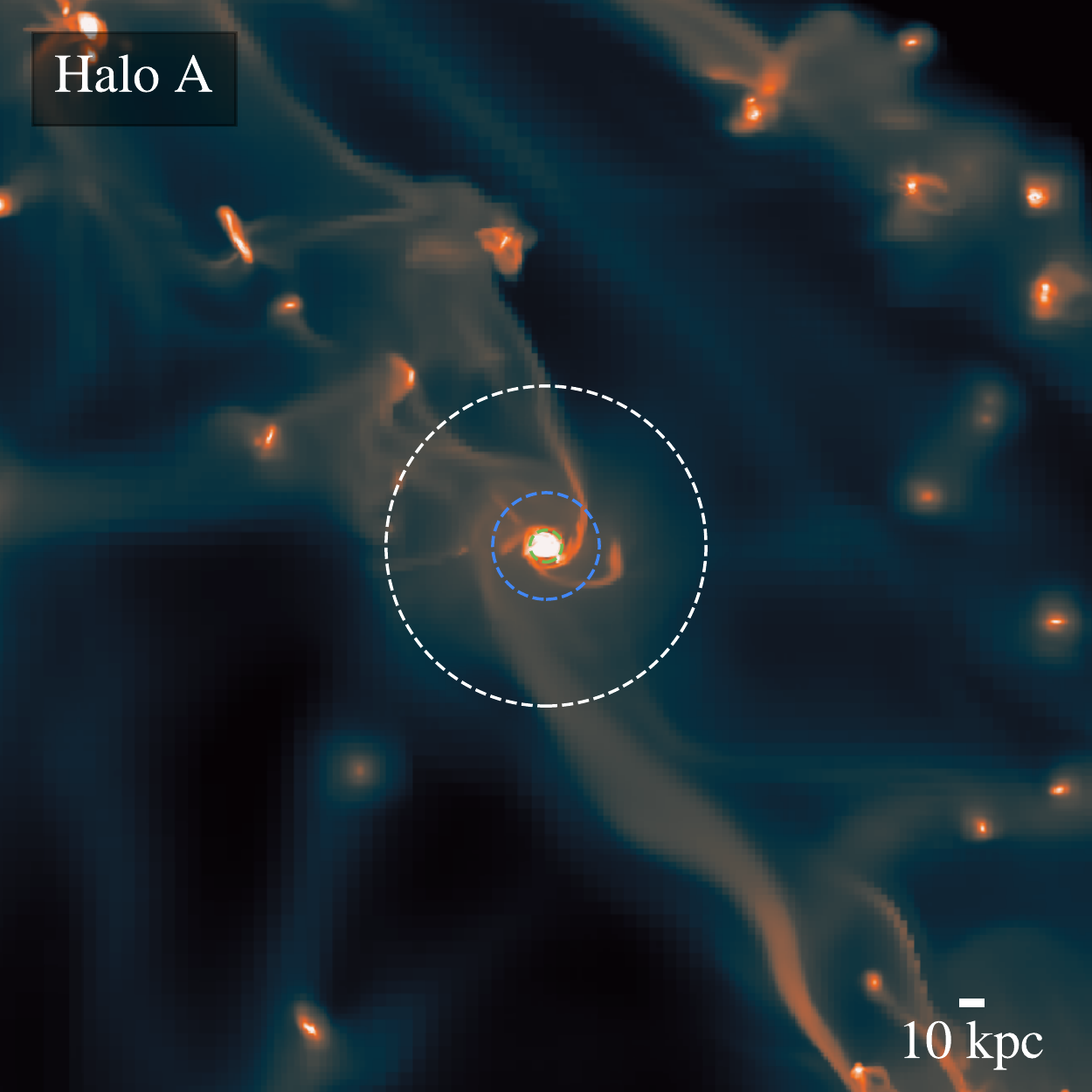}\hspace{0.1mm}%
        \includegraphics[height=2cm]{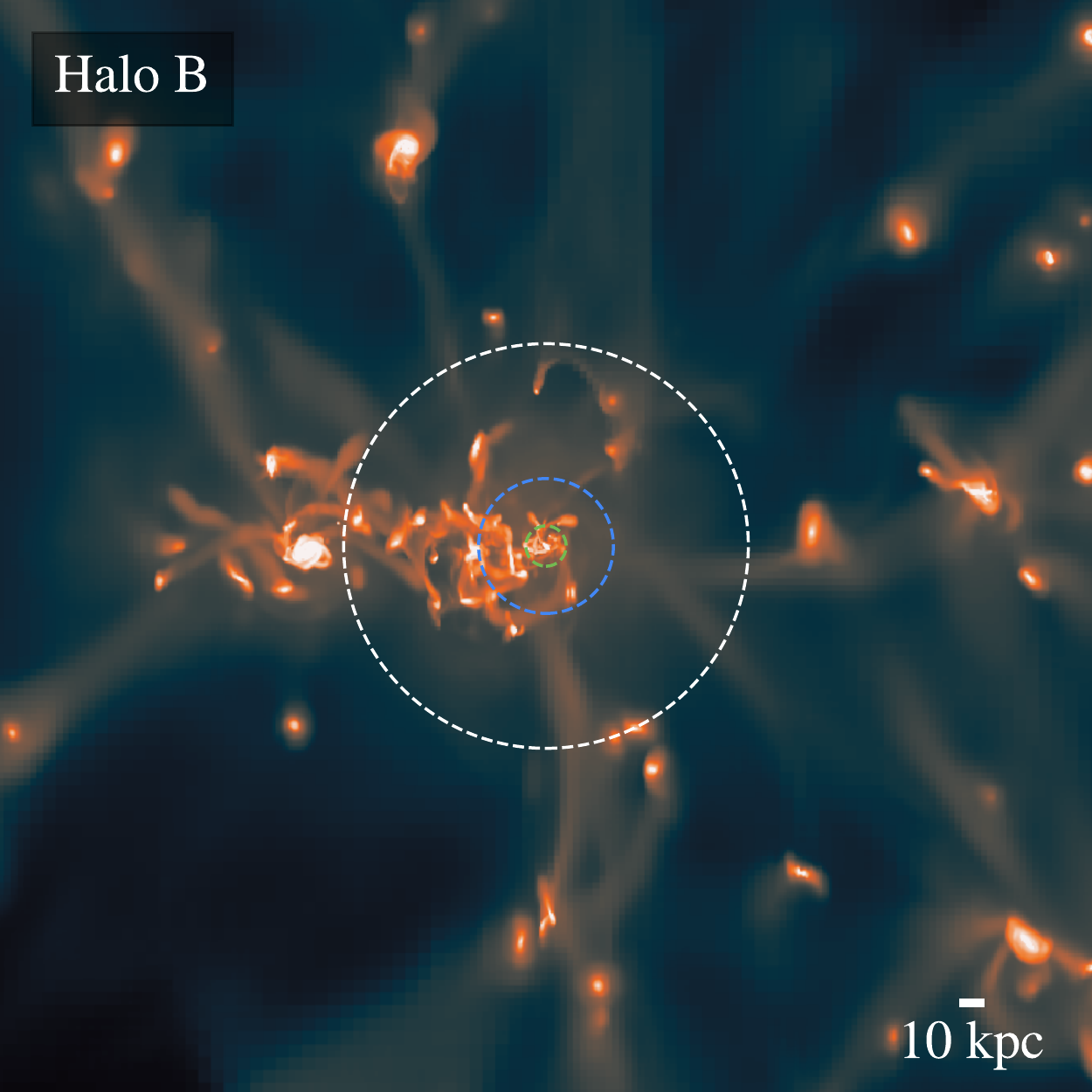}\hspace{0.1mm}%
        \includegraphics[height=2cm]{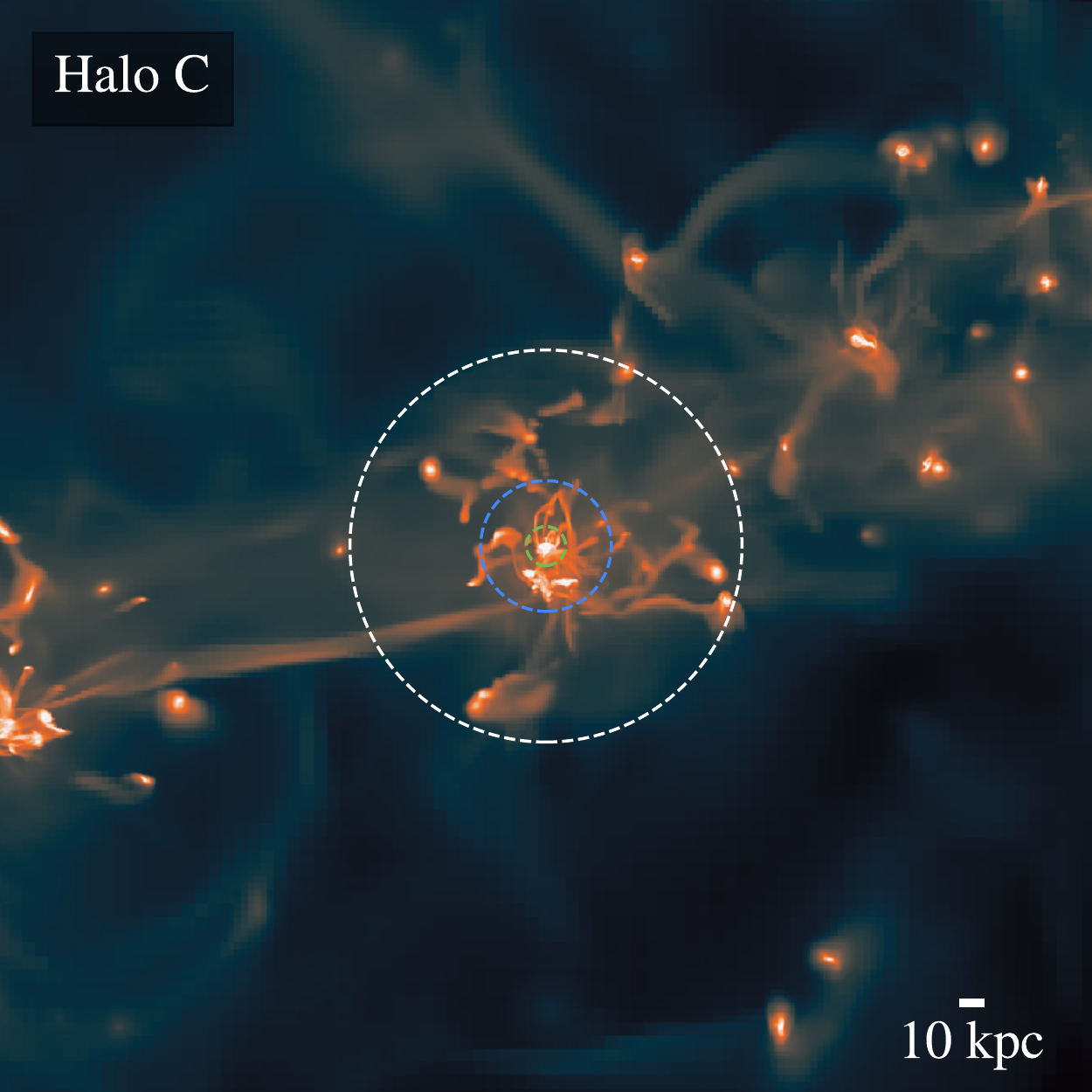}\hspace{0.1mm}%
        \includegraphics[height=2cm]{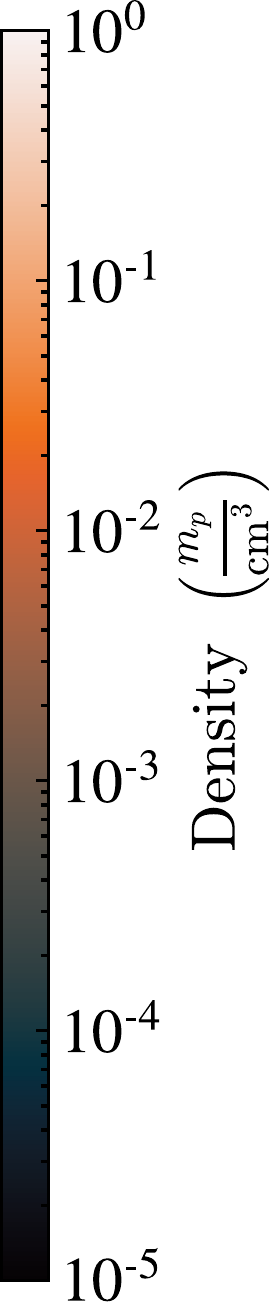}%
    }\vspace{-0.05cm}
    \resizebox{0.9\textwidth}{!}{
        \includegraphics[height=2cm]{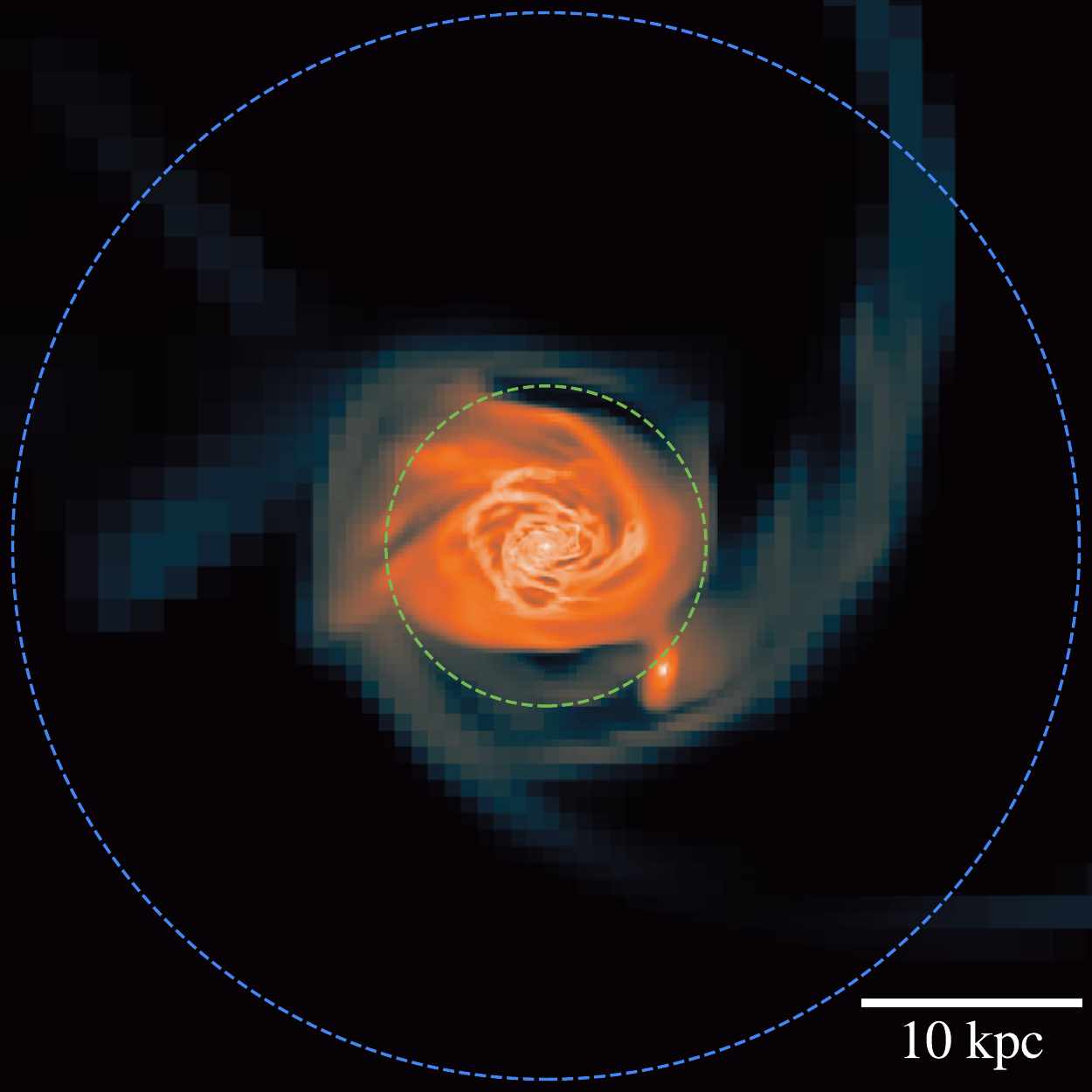}\hspace{0.1mm}%
        \includegraphics[height=2cm]{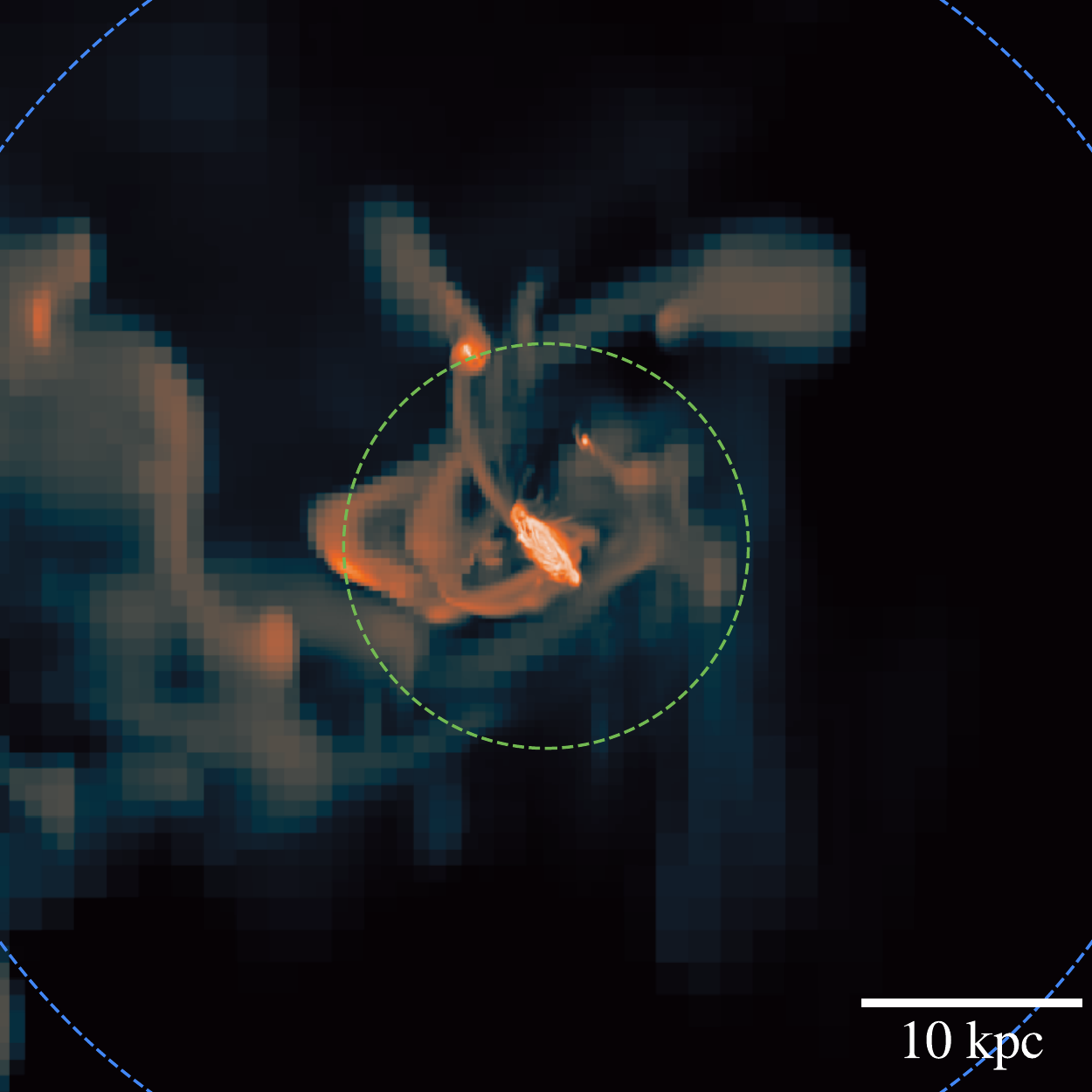}\hspace{0.1mm}%
        \includegraphics[height=2cm]{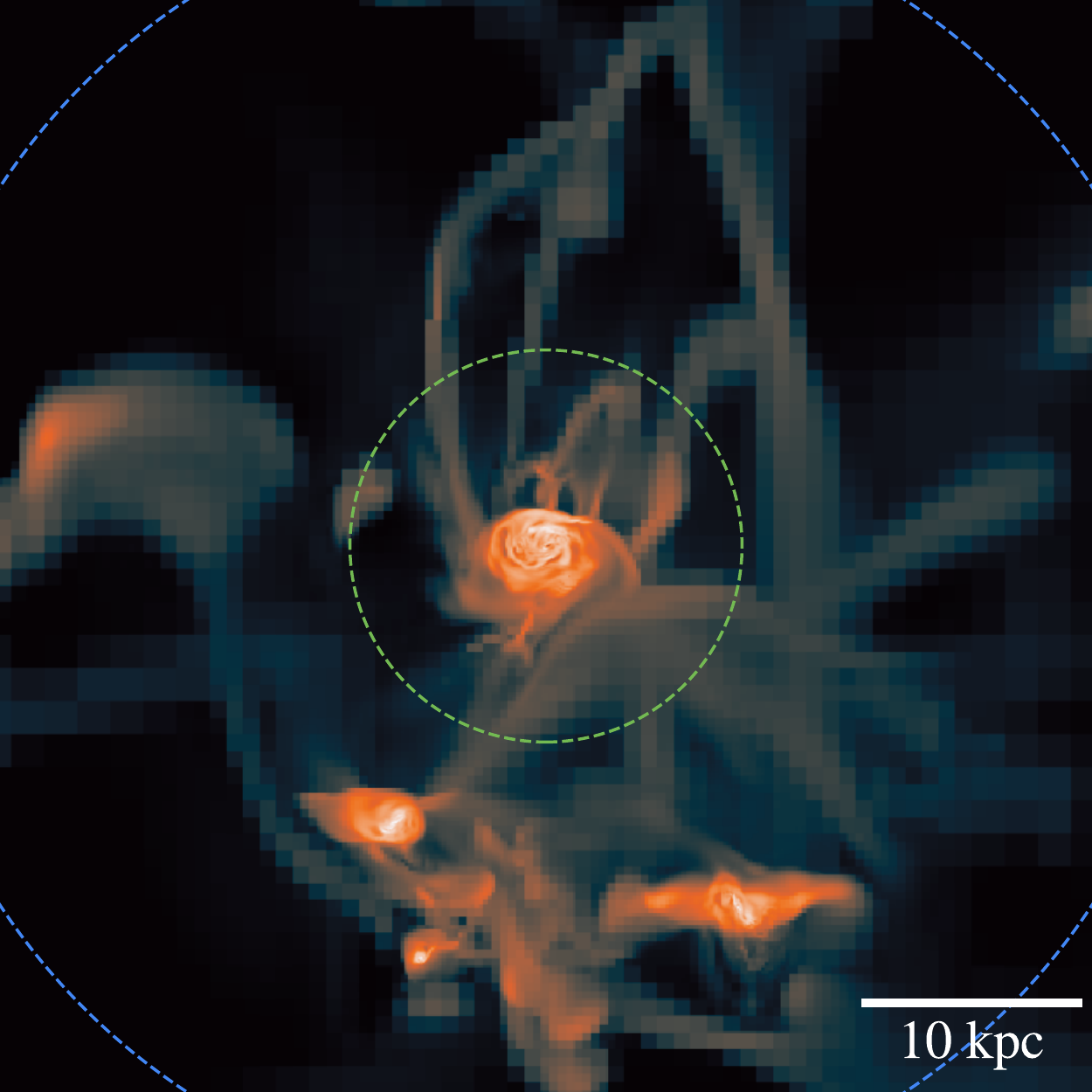}\hspace{0.1mm}%
        \includegraphics[height=2cm]{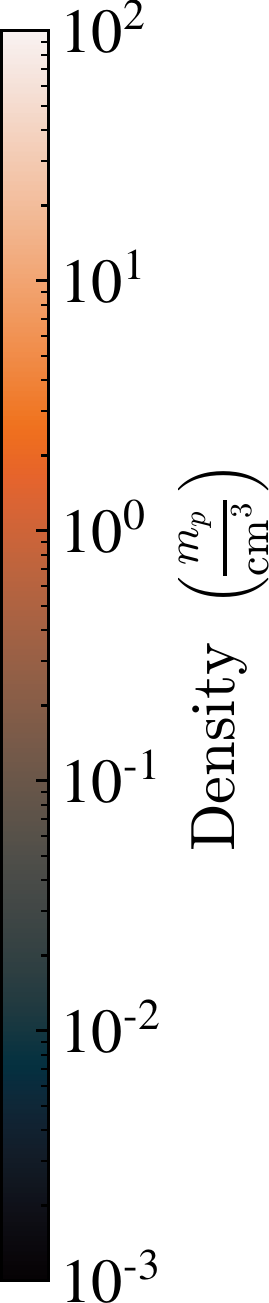}%
    }\vspace{0.1cm}
    \resizebox{0.9\textwidth}{!}{
        \includegraphics[height=2cm]{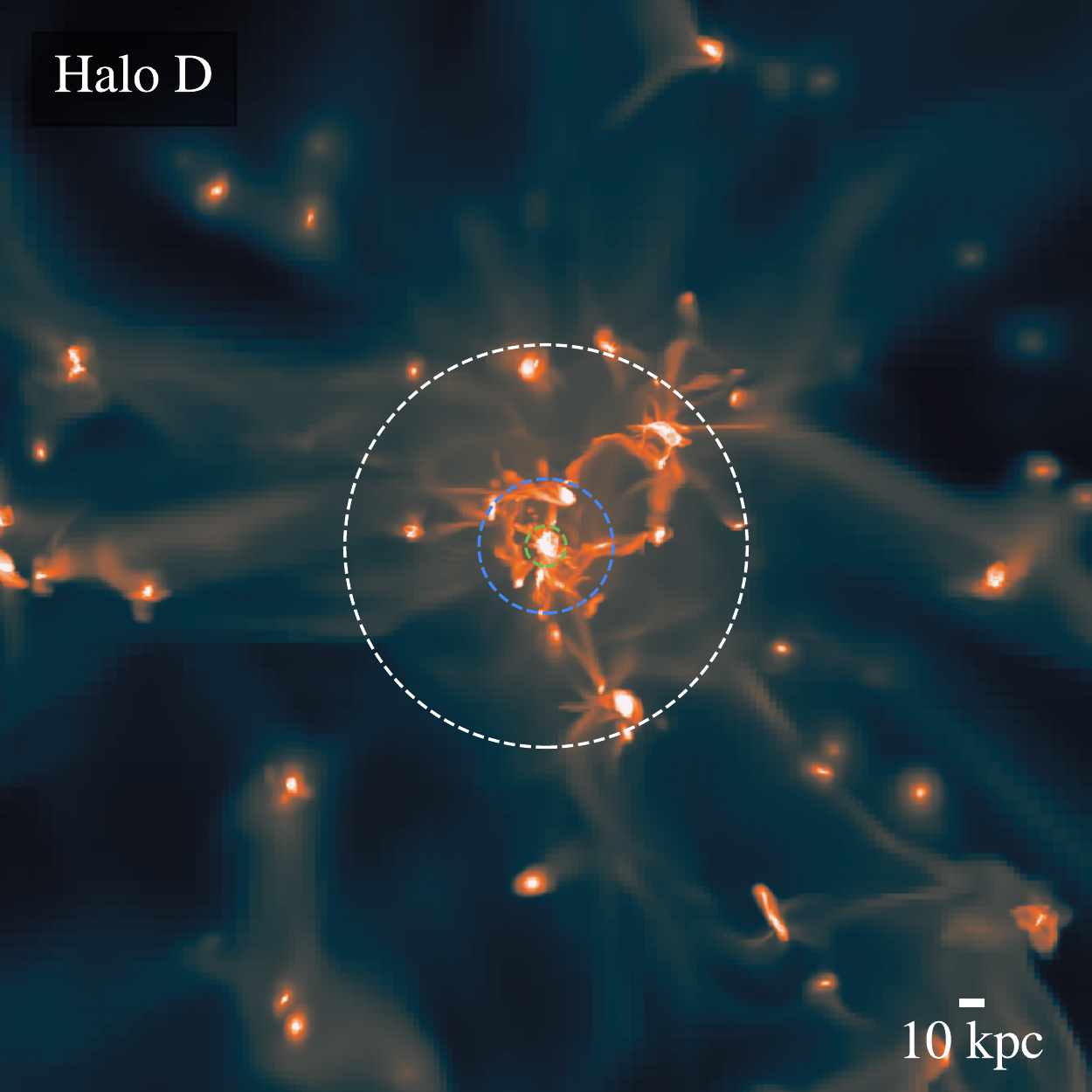}\hspace{0.1mm}%
        \includegraphics[height=2cm]{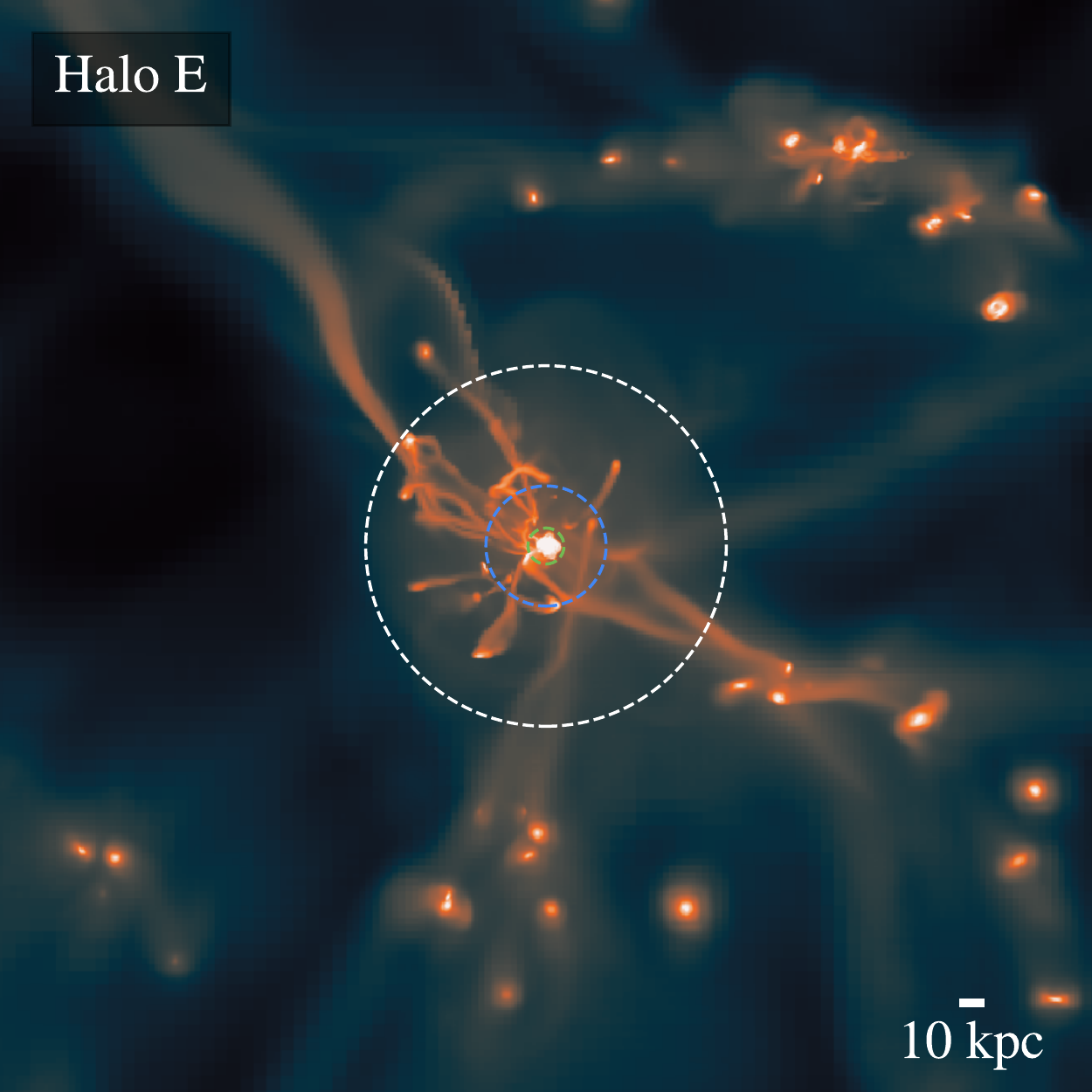}\hspace{0.1mm}%
        \includegraphics[height=2cm]{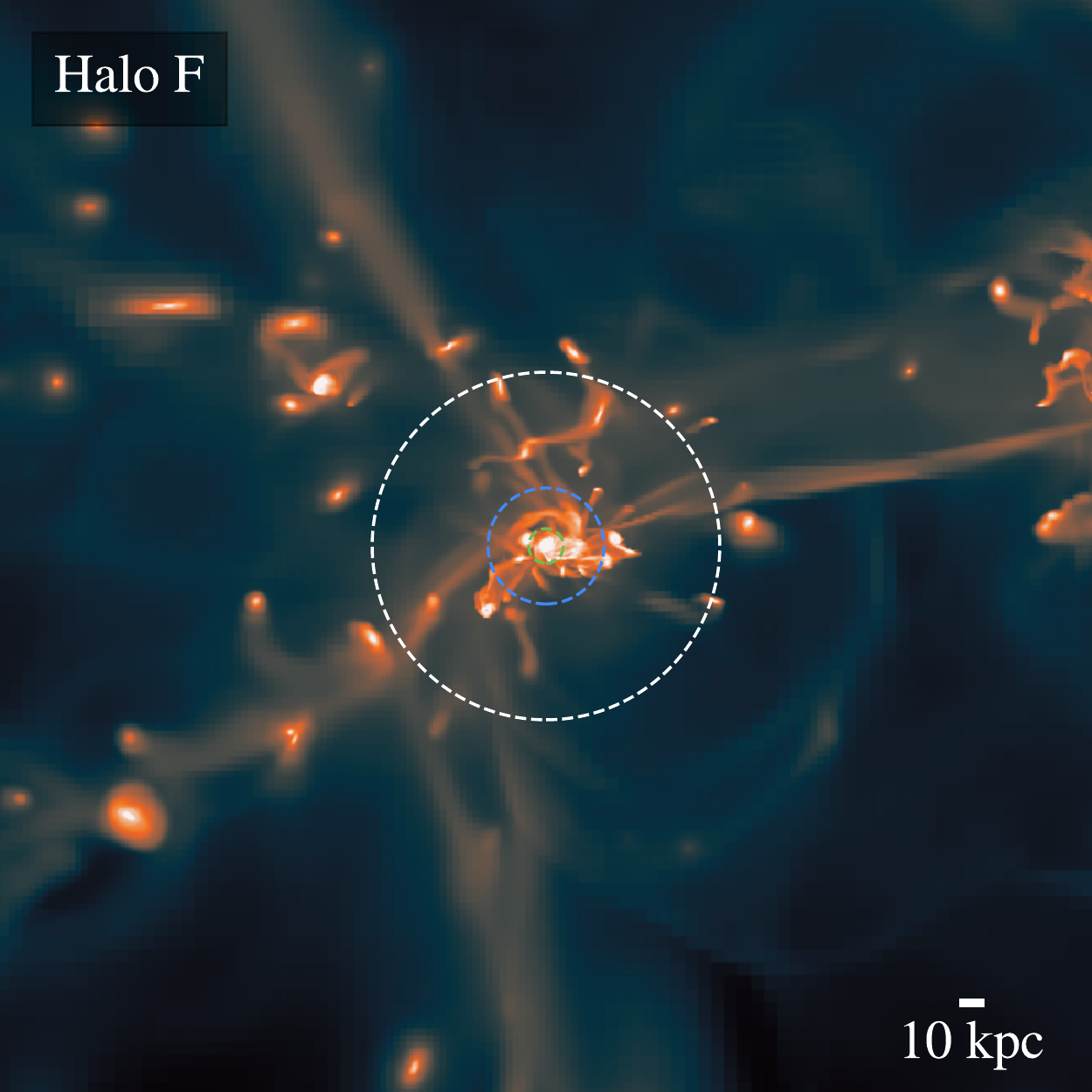}\hspace{0.1mm}%
        \includegraphics[height=2cm]{plots2/Projection_x_density_density_colorbar}%
    }\vspace{-0.05cm}
    \resizebox{0.9\textwidth}{!}{
        \includegraphics[height=2cm]{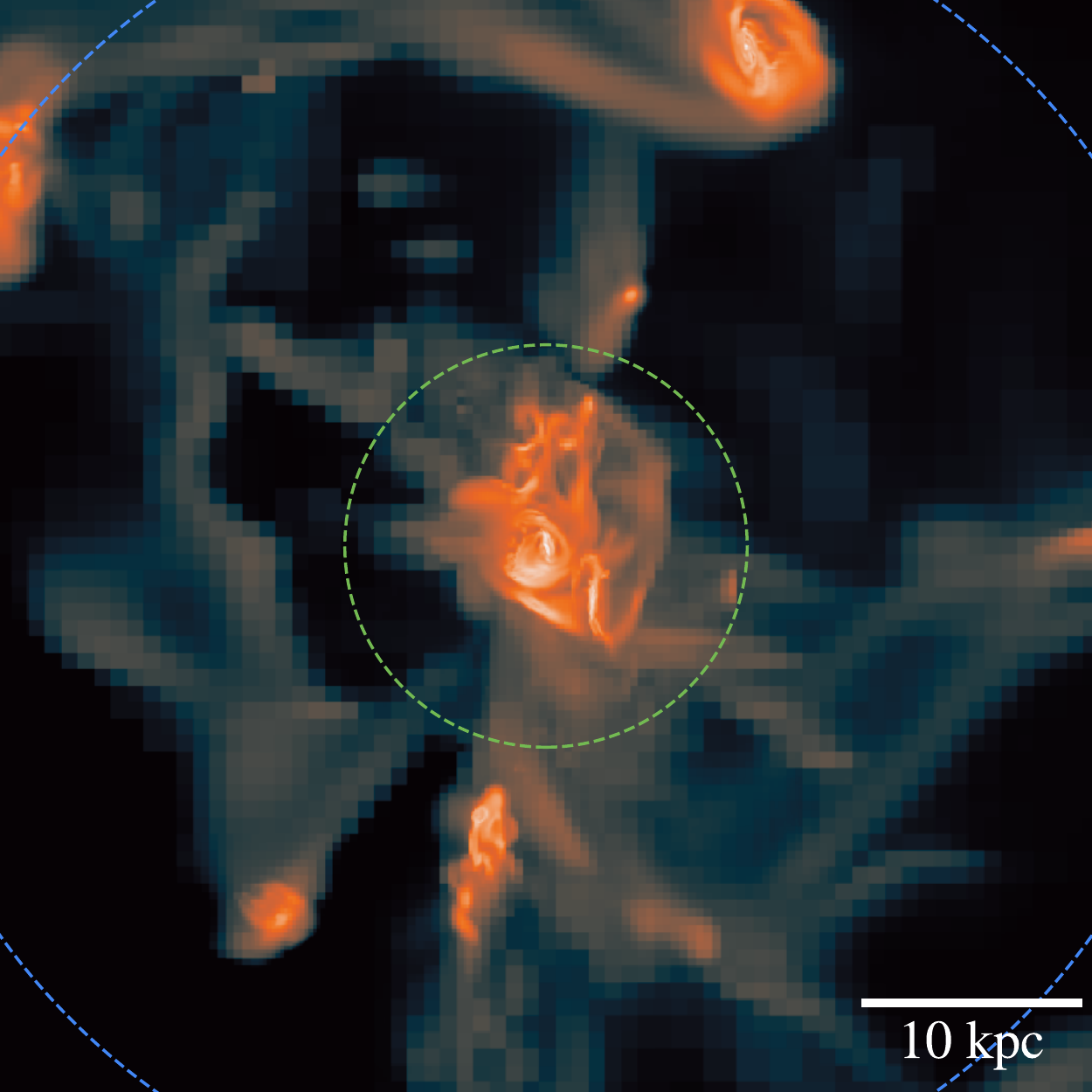}\hspace{0.1mm}%
        \includegraphics[height=2cm]{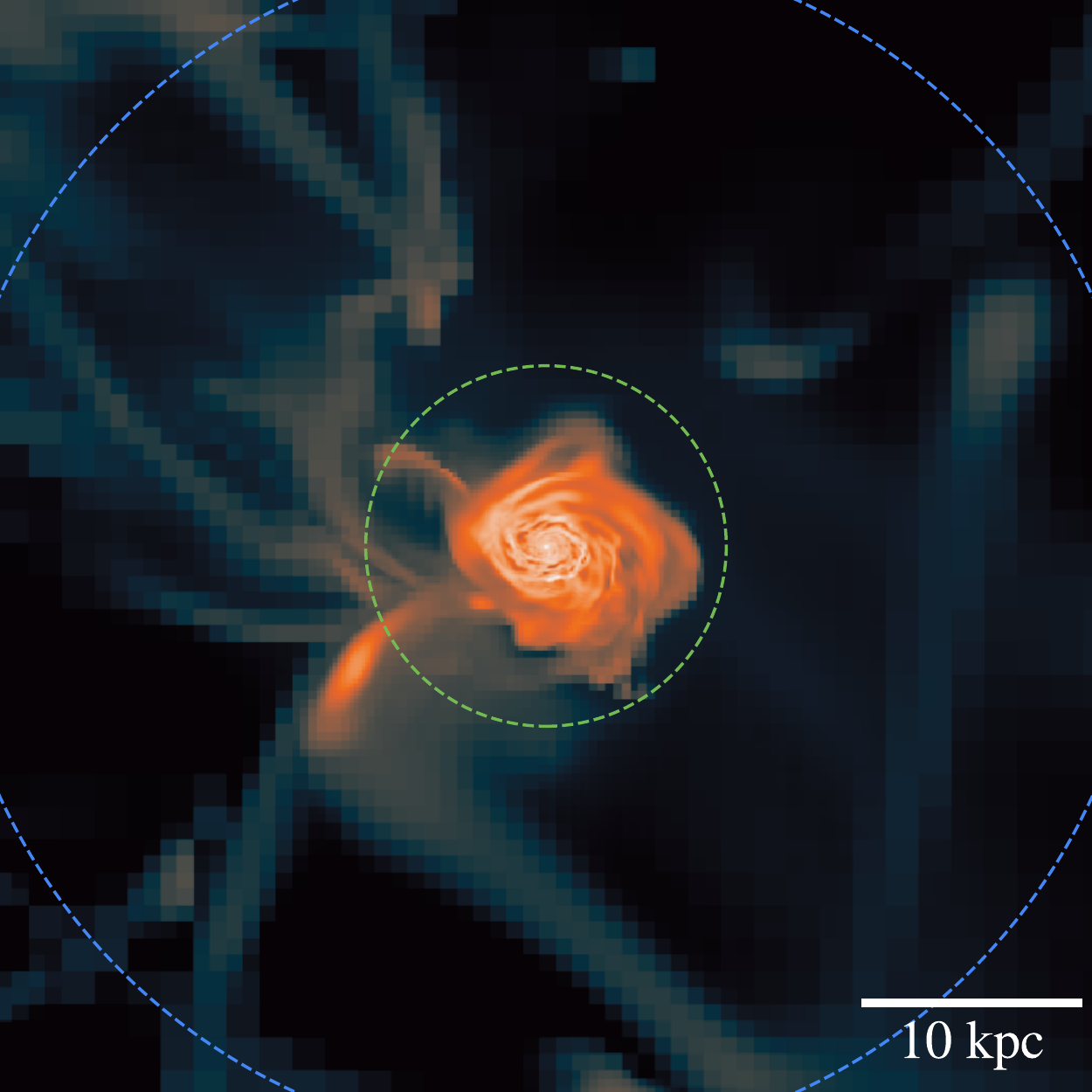}\hspace{0.1mm}%
        \includegraphics[height=2cm]{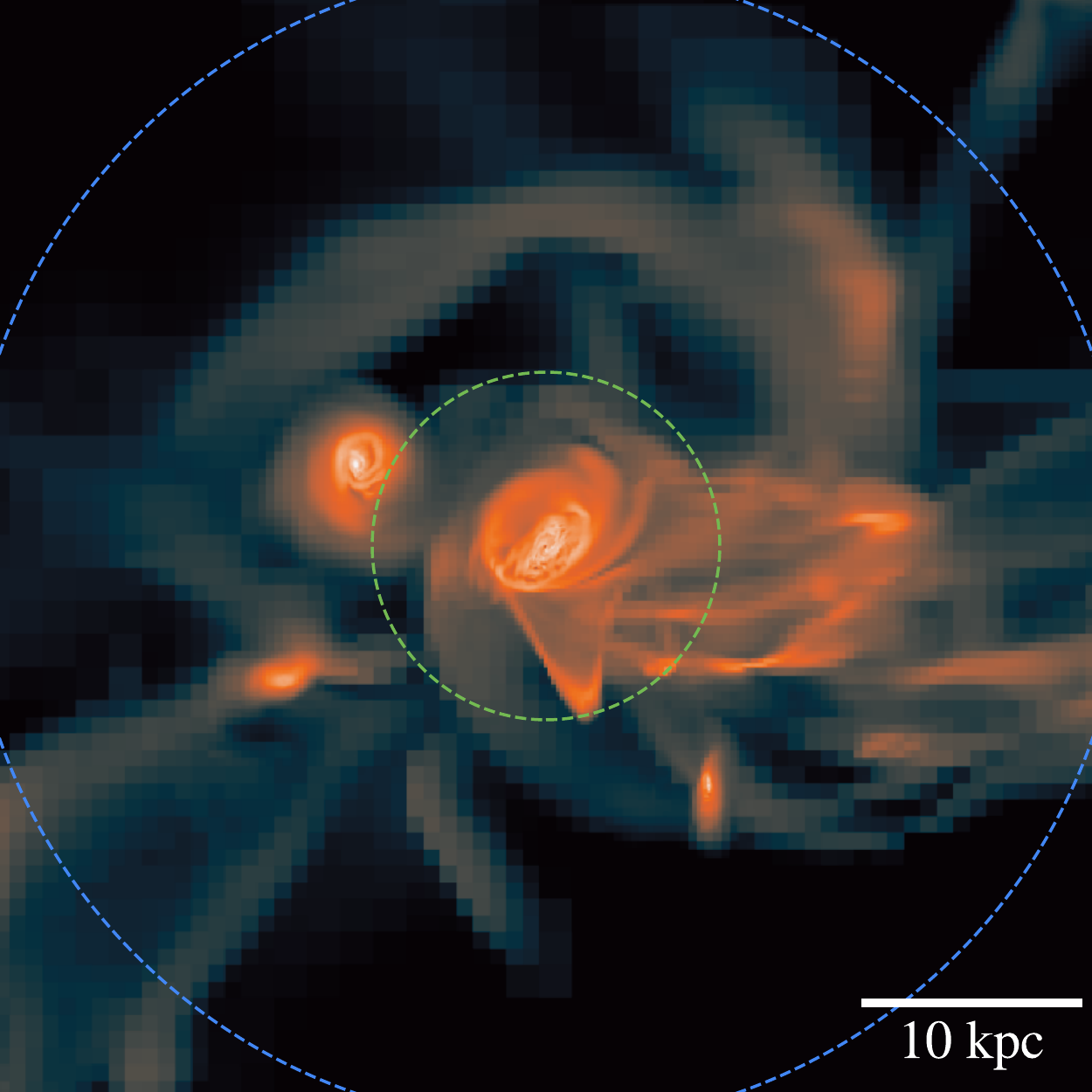}\hspace{0.1mm}%
        \includegraphics[height=2cm]{plots2/Projection_x_density_density_colorbar_zoom}%
    }
    \end{center}
    \caption{
        Density-weighted projections of the large-scale gas density (first and third rows) and the galaxy and CGM (second and fourth rows) of the six galaxies at $z=2$.
        The virial radius, $\Rvir$, $\Rvir/3$ and $\Rvir/10$ are shown as white, blue and green dashed circles respectively.
        They exhibit a complex large-scale structure made of filamentary cosmic streams and diffuse gas.
    }
    \label{fig:gas_and_star_at_z2}
\end{figure*}

\section{Introduction}

One of the successes of the $\Lambda$ cold dark matter ($\Lambda$CDM) model of the Universe is its ability to reproduce its large-scale structures observed in galaxy distribution \citep[\eg][]{Efstathiou_1981MNRAS,springel_large-scale_2006}.
These structures form out of the initial tiny density fluctuations of the primordial density field and under the effect of gravitational forces, matter departs from underdense regions to flow through cosmic sheets into filamentary structures.
Matter then flows from these filaments towards high-density peaks that will later become halos.
In the process, matter acquires kinetic properties \citep[\eg vorticity,][]{pichon_vorticity_1999,laigle_swirling_2015} in its journey through voids, sheets and filaments of the cosmic web, which, in turn, impact the assembly of dark matter halos.  %
Before shell crossing, baryons follow the same initial fate as dark matter (DM) and flow from underdense regions to sheets.
Yet, as they flow in sheets, pressure forces prevent them from shell-crossing so that they lose their normal velocity component to the shock front, dissipating their kinetic energy acquired at large-scale into internal energy (eventually radiated away by gas cooling processes).
Following potential wells created by dark matter, baryons also flow from sheets towards filamentary structures where they lose a second component of their velocity and reach a dense-enough state to efficiently cool radiatively, leading to a different structure in gas and DM density in filaments~\citep{pichon_rigging_2011,gheller2016,ramsoyetal21}.

At first order, galaxy formation is affected by the mass of their dark matter halo host and the local environment, as encoded by the local density on sub-Mpc scales, as it is assumed that baryons have the same past accretion history as dark matter \citep{1978MNRAS.183..341W,1998MNRAS.295..319M}.
These models have proven successful at explaining a number of observed trends, in particular against isotropic statistics, in the so-called halo model \citep[][]{2001MNRAS.323....1S}, yet they fail to explain some effects such as spin alignments \citep{tempel_galaxy_2013,codis_spin_2015,dubois_dancing_2014,chisari_galaxy-halo_2017}, colour  \citep{laigle_cosmos2015_2018,kraljic_galaxy_2018,kraljic_galaxies_2019} or star formation rates segregation \citep{malavasi_vimos_2017,kraljic_galaxies_2019,song2021}.
Indeed, the cosmic web sets preferred directions of accretion of cosmological gas that fuels star formation in the galaxy, and affects the orbital parameters of the successive mergers.
The detailed gas acquisition history, captured by the critical event theory developed in~\cite{cadiou_WhenCosmicPeaks_2020},  how much \AM is advected, as well as the origin of  mergers should thus all impact the formation of the galaxy.
Since the physical processes involved in dark matter halo formation differ from the baryonic processes operating during galaxy formation, one can expect that the cosmic web will have a specific impact on the formation of galaxies and may explain the disparity of their properties, such as morphology, in similar-looking dark matter halos.

In particular, at fixed halo mass and local density, properties of galaxies such as their colour or the kinematic structure vary with their location in the cosmic web.
One key process in the differential evolution of galaxies is gas accretion.
Indeed, at large redshifts, it has been suggested that the accretion of gas is dominated by flows of cold gas funnelled from the large scales to galactic scales \citep{birnboim_virial_2003,dekel_galaxy_2006}.
This mode of accretion has then been confirmed in numerical simulations using different methods \citep{keres_how_2005,dekel_galaxy_2006,ocvirk_bimodal_2008,nelson_moving_2013} as the source of a significant fraction of the baryonic mass but also \AM \citep{pichon_rigging_2011,kimm_angular_2011,stewart_angular_2013,stewart_high_2016}, which is decisive for several galactic properties~\citep{ceverinoetal10,agertzetal11,dekeletal2020,kretschmer_rapid_2020}. It has also been proposed that these flows may feed supermassive black holes \citep{di_matteo_cold_2012,dubois_feeding_2012}, which in turn affect the cold inflow rates \citep{dubois_blowing_2013}.

The cosmological origin of \AM is explained by tidal torque theory \citep[TTT,][]{peebles_origin_1969,white_AngularMomentumGrowth_1984,schafer_galactic_2009}. Though its predictions do not hold on a per-object basis \citep{porciani_TestingTidaltorqueTheory_2002a}, it can be extended to show that, indeed, \AM{} results from torques in the early Universe \citep{cadiou_AngularMomentumEvolution_2021}.
These torques in turn encode the anisotropy of the environments, which biases the \AM distribution to align it with the cosmic web \citep{codis_spin_2015}.
It is then expected that gas will fall in galaxies \emph{via} cold flows, feeding disks with angular-momentum rich gas that is itself aligned with the tides of the cosmic web.

Recent works have shown that the cold flows are subject to a variety of processes: they may fragment \citep{cornuault_are_2018} or be disrupted by hydrodynamical instabilities \citep{mandelker_instability_2016,mandelker_instability_2019}, but they are also sensitive to feedback events \citep{dubois_blowing_2013}.
In this context, \cite{danovich_four_2015} showed that in numerical simulations, cold flows are nevertheless able to feed galaxies with angular-momentum rich material \citep[as speculated by][]{pichon_rigging_2011,stewart_angular_2013}.
In this study, it was shown that the \AM acquired outside the halo at $z=2$ is transported down to the circumgalactic medium (CGM); the gas then settles in a ring surrounding the disk, where gravitational torques spin the gas down to the mean spin of the baryons.
Another study, albeit at larger redshifts, found that the dominant force was pressure \citep{prieto_how_2017}.
Since there is not much freedom on the final \AM of the galaxies, as constrained by their radius, the excess \AM brought by cold flows
has to be redistributed somehow before it reaches the disk.

The details of where this \AM will end up are key to understanding the \AM distribution in galaxies, but also to understanding to what extent their spin is aligned with the cosmic web.
If the dominant torques acting on the \AM are pressure torques, resulting from internal processes (SN winds, AGN feedback bubbles and hydrodynamical shocks), then the spin of the galaxy would likely be a result of chaotic internal processes and would lose its connection to the cosmic web.
On the contrary, if gravitational torques dominate, then the spin-down of the cold gas is likely to drive a spin-up of either the disk or the dark matter halo, which themselves are the result of their past \AM accretion history.
In this last scenario, the details of which part(s) of the halo or the disk interact and exchange \AM with the infalling material would constrain models aimed to understand the evolution of the spin of galaxies.

Historically, the study of cold accretion has been particularly challenging in numerical simulations.
Early simulations using Smooth Particle Hydrodynamics (SPH) methods largely overestimated the fraction of gas accreted cold \citep[see \eg][for a discussion on this particular issue]{nelson_moving_2013} as a result of the difficulty to capture shocks using SPH.
Adaptive Mesh Refinement (AMR) simulations do not suffer from this caveat \citep{ocvirk_bimodal_2008}, yet they fail at providing the Lagrangian history of the gas -- in particular its past temperature -- which is required to detect the cold-accreted gas.
In order to circumvent this limitation, most simulations relied on velocity-advected tracer particles \citep{dubois_blowing_2013,tillson_angular_2015}.
However, this approach yields a very biased tracer distribution that fails at reproducing correctly the spatial distribution of gas in filaments: most tracer particles end up in convergent regions (centre of galaxies, centre of filaments) while divergent regions are under-sampled.
In order to reproduce more accurately the gas distribution, \cite{genel_following_2013} suggested relying on a Monte-Carlo approach where tracer particles follow mass fluxes instead of being advected.
This was later improved to include the entire Lagrangian evolution of the gas through star formation and feedback, as well as AGN accretion and feedback \citep{cadiou_AccurateTracerParticles_2019}.
The aim of this paper is to investigate the evolution of the \AM of the cold and hot gas using cosmological simulations of group progenitors at $z>2$.
We provide a detailed study of the evolution of the \AM of the cold and hot gas.
In particular, the question of which forces are responsible for the spin-down and realignment of the \AM of the gas accreted in the two modes of accretion (hot and cold) will be addressed.

\Cref{sec:numerical.method} presents the numerical setup.
\Cref{sec:numerical.results} presents the \AM evolution of the cold and hot gas. It follows the evolution of the magnitude and orientation of the \AM and the different forces and torques at play in the different regions of the halos.
It details the evolution of the magnitude and orientation of the \AM and the different forces and torques at play in the different regions of the halos.
In \Cref{sec:numerical.discussion}, we present their implication on the distribution of \AM in the galaxy and the CGM.
Finally, \cref{sec:numerical.conclusion} wraps things up and concludes.

In the following, we will adopt a similar naming convention as \cite{danovich_four_2015}.
We will write $\Rvir$ the virial radius of a halo. The outer halo is defined as the region between $\Rvir$ and $\Rvir/3$.
The circumgalactic medium (CGM) is defined as the region between $\Rvir/3$ and $\Rvir/10$.
The `disk' is the region at radius $r<\Rvir/10$ where the galaxy is found.
\section{Method}
\label{sec:numerical.method}

We describe the numerical setup in \cref{sec:numerical.desc_simu}. We then detail the equations driving the \AM evolution in \cref{sec:numerical.equations}, and the details of the extraction of the different torques in \cref{sec:numerical.methods.torque-extraction}.
In \cref{sec:numerical.cold-gas-selection}, we describe how we selected the cold gas being accreted on the halos in the simulations.

\subsection{Simulations}
\label{sec:numerical.desc_simu}

We produced a suite of three zoomed-in regions in a \SI{50}{h^{-1}\, cMpc}-wide cosmological box, hereafter named S1, S2, S3.
The three simulations contain six high-resolution halos with virial mass $4\times10^{11}\lesssim M_{\rm vir}/{\rm \Msun}\lesssim 7\times10^{11}$ at $z=2$, hereafter named A, B, C, D, E and F (see their properties in \cref{table:halo-props}), containing only high-resolution dark matter (DM) particles within twice the virial radius of the halo at $z=2$.
We adopted a cosmology that has a total matter density of $\Omega_\mathrm{m} = 0.3089$, a dark energy density of $\Omega_\Lambda = 0.6911$, a baryonic mass density of $\Omega_\mathrm{b}= 0.0486$, a Hubble constant of $H_0 = \SI{67.74}{\km.s^{-1}.Mpc^{-1}}$, a variance at \SI{8}{Mpc} $\sigma_8 = 0.8159$, and a non-linear power spectrum index of $n_s = 0.9667$, compatible with a Planck 2015 cosmology~\citep{planck_collaboration_planck_2015}.
We generated the initial conditions with \textsc{MUSIC}~\citep{hahnabel11}.
The simulations are started with a coarse grid of $128^3$ (level 7) and several nested grids with increasing levels of refinement up to level 11, corresponding to a DM mass resolution of respectively $6.3\times 10^9\, \rm M_\odot$ and $1.5\times 10^6\, \rm M_\odot$.

\begin{table}
    \centering
    \caption{Name of the halo, name of the simulation, virial mass of the halo and the stellar mass of their central galaxy at $z=2$.}
    \begin{tabular}{llccc}
        \toprule
        Name     & Simulation & $M_\mathrm{vir}/{10^{11}}\ \si{\Msun}$ & $M_\star/10^{10}\ \si{\Msun}$\\
        \midrule
        A         & S1  & \num{3.66} & \num{6.07} & \\
        B         & S2  & \num{7.82} & \num{9.20} & \\
        C         & S3  & \num{6.64} & \num{5.09} & \\
        D         & S1  & \num{7.29} & \num{4.18} & \\
        E         & S1  & \num{5.23} & \num{7.84} & \\
        F         & S3  & \num{4.63} & \num{3.49} &\\
        \bottomrule
    \end{tabular}
    \label{table:halo-props}
\end{table}

\begin{figure}
    \includegraphics[width=\columnwidth,trim={0 4cm 0 4cm}, clip]{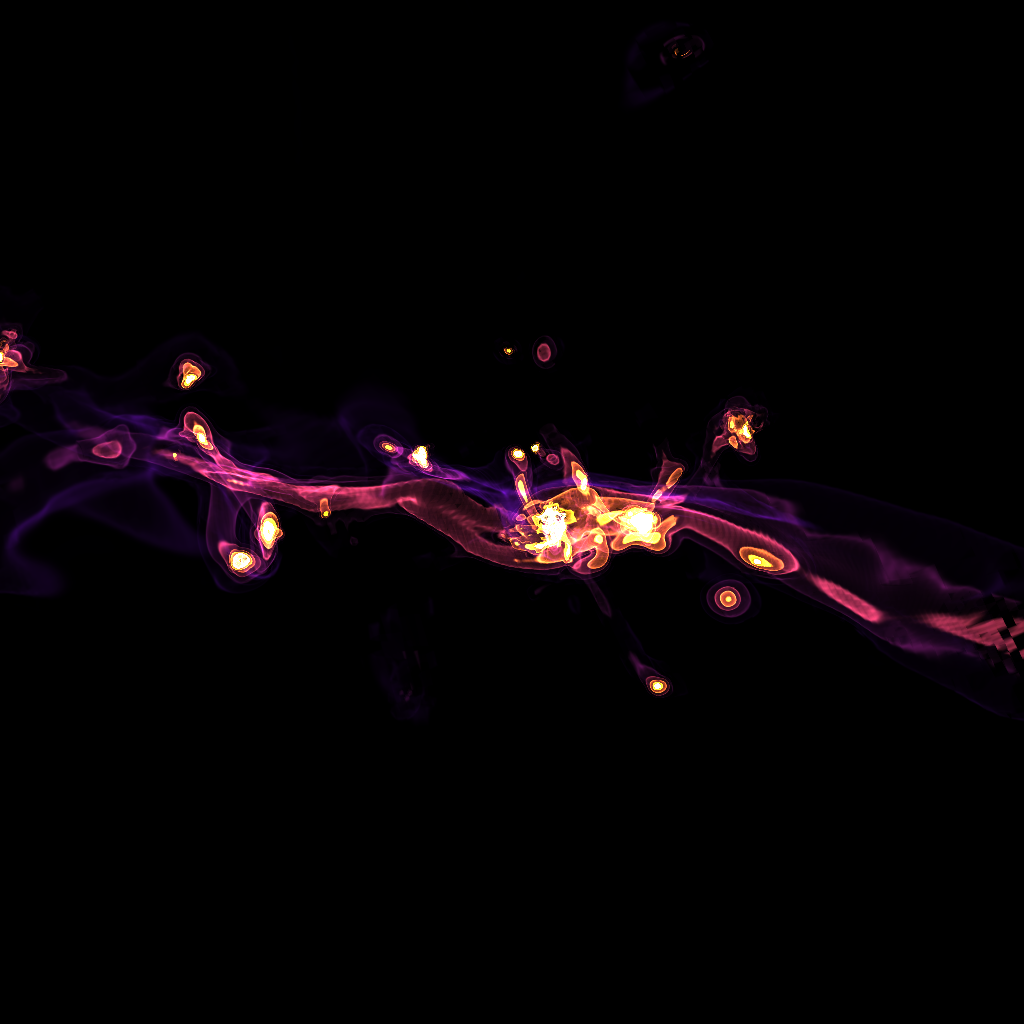}
    \caption{
        Volume rendering of the density isocontours of the cold gas around galaxy~A at $z=3.6$.
        The cold gas is found in infalling satellites, while the filamentary structure, which connects the large-scale structure filamentary structure to the galaxy (at the centre), is obvious.
    }
    \label{fig:3drendering}
\end{figure}

The halos are simulated with the adaptive mesh refinement code \textsc{ramses}~\citep{Teyssier02}.
Particles (dark matter, stars, black holes) are moved with a leap-frog scheme, and to compute their contribution to the gravitational potential, their mass is projected onto the mesh with a cloud-in-cell interpolation.
Gravitational acceleration is obtained by computing the gravitational potential through the Poisson equation numerically obtained with a conjugate gradient solver on levels above 12, and a multigrid scheme~\citep{guillet_multigrid_2011} otherwise.
We include hydrodynamics in the simulations, which system of non-linear conservation laws is solved with the MUSCL-Hancock scheme~\citep{vanleer77} using a linear reconstruction of the conservative variables at cell interfaces with minmod total variation diminishing scheme, and with the use of the HLLC approximate Riemann solver~\citep{toro} to predict the upstream Godunov flux.
We allow the mesh to be refined according to a quasi-Lagrangian criterion: if $\rho_{\rm DM}+\rho_{\rm b}/f_{\rm b/DM}>8 m_{\rm DM,res}/\Delta x^3$, where $\rho_{\rm DM}$, and $\rho_{\rm b}$ are respectively the DM and baryon density (including stars, gas, and supermassive black holes (SMBHs)), and where $f_{\rm b/DM}$ is the universal baryon-to-DM mass ratio.
Conversely, an oct (8 cells) is derefined when this local criterion is not fulfilled.
The maximum level of refinement is also enforced up to 4 minimum cell size distance around all SMBHs.
The simulations have a roughly constant proper resolution of \SI{35}{pc} (one additional maximum level of refinement at expansion factor $0.1$ and $0.2$ corresponding to maximum level of refinement of respectively 18 and 19), a star particle mass resolution of $m_{\star,\rm res}=\SI{1.1e4}{\Msun}$, and a gas mass resolution of \SI{2.2e5}{\Msun} in the refined region.
We employ Monte-Carlo tracer particles to sample mass transfers between cells and the various baryonic components, as described in~\cite{cadiou_AccurateTracerParticles_2019}. Each tracer samples a mass of $m_\mathrm{t} = \SI{2.0e4}{\Msun}$ ($N_\mathrm{tot}\approx\num{1.3e8}$ particles).
There is on average 0.55 tracer per star and 22 per initial gas resolution element.
Cells of size \SI{35}{pc} and gas density of \SI{20}{cm^{-3}} contain on average one tracer per cell.

We also add additional baryonic physics relevant to the process of galaxy formation at \SI{35}{pc} similar to the physics of NewHorizon~\citep{duboisetal20}.
The simulations include a metal-dependent tabulated gas cooling function following \cite{sutherland_cooling_1993} for gas with temperature above $T> \SI{e4}{K}$.
The metallicity of the gas in the simulation is initialised to $Z_0 = \SI{e-3}{Z_\odot}$ to allow further cooling below \SI{e4}{K} down to $T_\mathrm{min} = \SI{10}{K}$ \citep{rosen_global_1995}.
Reionisation occurs at $z=8.5$ using the \cite{haardt_radiative_1996} UV background model and assuming gas self-shielding above \SI{e-2}{H.cm^{-3}}.
Star formation is allowed above a gas number density of $n_0=\SI{10}{H.cm^{-3}}$ with a Schmidt law, and with an efficiency $\epsilon_{\rm ff}$ that depends on the gravo-turbulent properties of the gas \citep[for a comparison with a constant efficiency see][]{nunez2020}.
The stellar population is sampled with a \cite{kroupa_variation_2001} initial mass function, where $\eta_{\rm SN} = 0.317$ and the yield (in terms of mass fraction released into metals) is $0.05$.
Type II supernovae are modelled with the mechanical feedback model of~\cite{kimm_towards_2015} with a boost in momentum due to early UV pre-heating of the gas following \cite{geen_detailed_2015}.
The simulations also track the formation of SMBHs and their energy release through AGN feedback.
SMBH accretion assumes an Eddington-limited Bondi-Hoyle-Littleton accretion rate in jet mode (radio mode) and thermal mode (quasar mode) using the model of \cite{dubois_feeding_2012}.
The jet is modelled self-consistently by following the \AM of the accreted material and the spin of the black hole \citep{dubois_black_2014}.
The radiative efficiency and spin-up rate of the SMBH are computed assuming the radiatively efficient thin accretion disk from~\cite{shakura&sunyaev73} for the quasar mode, while the feedback efficiency and spin-up rate in the radio mode follow the prediction of the magnetically choked accretion flow model for accretion disks from~\cite{mckinney_general_2012}.
SMBHs are created with a seed mass of \SI{e4}{\Msun} for S1 and \SI{e5}{\Msun} for S2 and S3.
For the exact details of the spin-dependent SMBH accretion and AGN feedback, see~\cite{duboisetal20}.
Compared to NewHorizon, the main difference is that we do not boost the dynamical friction of the gas onto SMBHs.
This likely result in our AGN feedback being weaker and may explain why our simulated galaxies have larger stellar-to-halo than what one would expect from abundance matching (see \cref{table:halo-props}).
A detailed investigation of the origin of this issue is however beyond the scope of the paper.

We extract halo catalogues using \adaptahop{} \citep{aubert_OriginImplicationsDark_2004}, using the `Most massive Sub-node Method' and the parameters proposed in \cite{tweed_building_2009}. The density is computed from the 20 nearest neighbours and we use a linking length parameter of $b=0.2$.
Figure~\ref{fig:gas_and_star_at_z2} shows the projected gas density at $z=2$ of the six simulated halos with their virial radius on top of each image as identified by the halo finder.

\subsection{Evolution of the specific angular momentum}
\label{sec:numerical.equations}
The equation driving the evolution of the \sAM of the gas $\vec{l} = \rr \cross \vv$, where $\rr$ and $\vv$ are respectively the distance and velocity of the gas with respect to the halo centre, is obtained from the equation of motion
\begin{align}
    \pdv{\vv}{t} + (\vv\vdot\vec{\nabla}) \vv &= -\frac{\grad P}{\rho} - \grad\phi.
    \label{eq:euler}
\end{align}
Taking the time derivative of the \sAM, we obtain that
\begin{equation}
    \dv{\vec{l}}{t} = \rr\cross\left(\pdv{\vv}{t} + (\vv\vdot\vec{\nabla})\vv \right) + \left(\pdv{\rr}{t} + (\vv\vdot\vec{\nabla})\rr \right) \cross\vv.
    \label{eq:l-evolution-eq}
\end{equation}
After trivial algebra, the rightmost part of the right-hand side vanishes. Using \cref{eq:euler,eq:l-evolution-eq}, the Lagrangian time derivative of the \sAM then reads
\begin{equation}
    \dv{\vec{l}}{t} = \vec{\tau}_{P} + \vec{\tau}_\phi,
    \label{eq:sAM-evolution}
\end{equation}
where $\vec{\tau}_P = -\vec{r}\cross \grad P / \rho$, $\vec{\tau}_\phi = -\vec{r}\cross\grad \phi$ are the specific pressure and gravitational torques. Here $P$ and $\rho$ are the pressure and density of the gas and $\phi$ is the gravitational potential.
The potential is obtained from Poisson equation
\begin{equation}
    \laplacian\phi = 4\pi G\rho_\mathrm{tot},
    \label{eq:poisson}
\end{equation}
where $\rho_\mathrm{tot}$ is the {total} matter density (DM, stars, gas and SMBHs) and $G$ is the gravitational constant.

\subsection{Torque extraction}
\label{sec:numerical.methods.torque-extraction}

Most of the previous works \citep{danovich_four_2015,prieto_how_2017} have studied the relative contribution of each torque to the \AM evolution of the cold gas focusing in particular on their magnitude and direction, splitting the torques between pressure and gravitational torques.
In this section, we provide an improvement over these past works by computing the gravitational torques from each source (stars, DM and the gas) separately.
We also lay down a general method to compute gradients in post-processing in AMR codes, which we then use to compute pressure gradients, and in particular, pressure torques.

\begin{figure}
    \centering
    \includegraphics[width=\columnwidth]{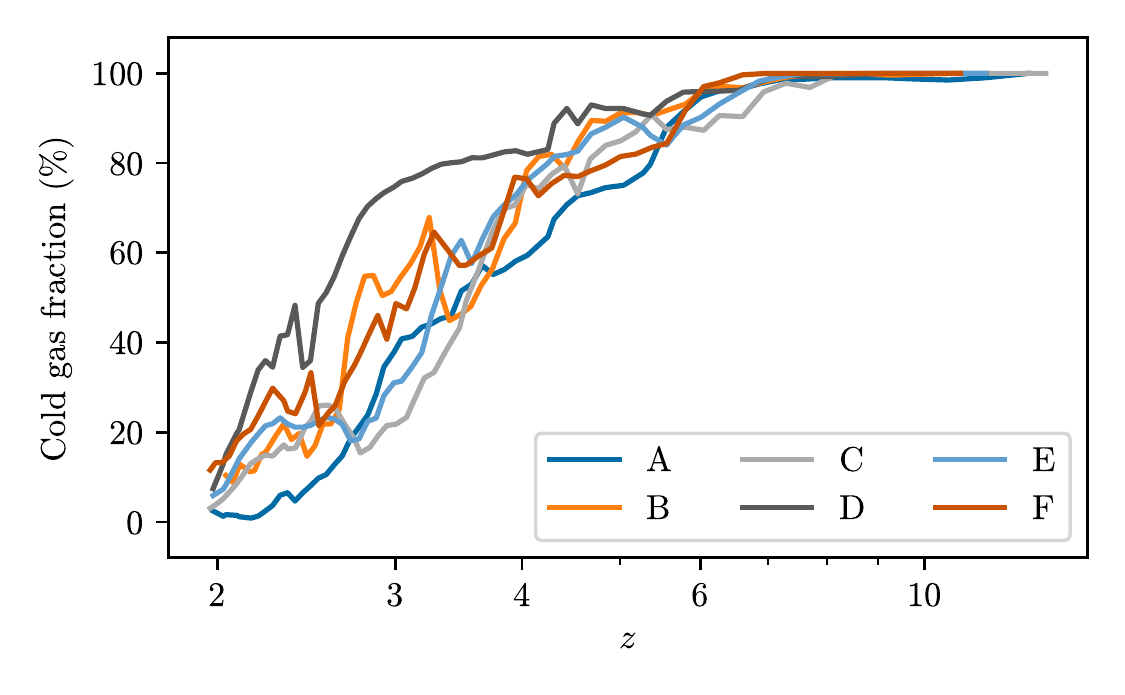}
    \caption{
        Evolution of the fraction of gas accreted cold as a function of time for all six galaxies.
        The fraction is evaluated using the Lagrangian history of the tracer particles.
        A tracer particle is considered accreted at $z$ if it crossed $R_\mathrm{vir}(z)$ inwards for the first time.
        It is considered cold if it never heated above \SI{2.5e5}{K} between $2$ and $0.3 R_\mathrm{vir}$.
        The cold gas fraction decreases after $z=6$, which coincides with the time at which the halo mass reaches \SI{e11}{\Msun}.
    }
    \label{fig:cold_gas_fraction}
\end{figure}

\subsubsection{Gravitational torques}
\label{sec:numerical.methods.grav_torques}

In the vicinity of galaxies, the different massive sources (DM, stars, gas\footnote{Here we ignore SMBHs, as their gravitational potential may only dominate very close to the center of the galaxy.}) all contribute to the total gravitational potential $\phi = \phi_\mathrm{DM} + \phi_\star + \phi_\mathrm{gas}$ \emph{via} their own Poisson equation
\begin{equation}
    \laplacian \phi_i = 4\pi G \rho_i,
    \label{eq:poisson-explicit}
\end{equation}
where $\phi_i$ and $\rho_i$ are the gravitational potential and the density of the component $i$ (DM, stars, gas).
One can then compute the specific forces resulting from each potential $\vec{F}_i = -\grad \phi_i$ which can then be used to compute the specific torques at position $\vec{r}$
\begin{equation}
    \vec{\tau}_i \equiv \vec{r} \cross \vec{F}_i.
\end{equation}
In order to extract the torques resulting from each gravitational source, we have modified the code \ramses{} to extract in post-processing the specific forces due to the different matter components (DM, gas, stars).
This was performed by stripping down \ramses{} to keep only the Poisson solver, applied to the density of each individual component\footnote{The fiducial implementation solves the Poisson equation directly on the total matter density (gas + stars + DM).}.
\subsubsection{Pressure torques}
\label{sec:numerical.method.pressure_gradients}

The precise capture of shocks is fundamental to most astrophysical codes.
These shocks then result in strong gradients which are usually captured by a few cells with a MUSCL-Hancock scheme.
While numerical codes routinely deal with strong gradients, most AMR post-processing tools either do not provide any utility to compute them (pynbody, \citealp{pontzen_pynbody_2013}; pymses, \citealp{guillet_pymses_2013}), or have gradient computing capacities that are not available for octree-based AMR datasets, as is the case with \ramses{} \citep[\eg yt,][]{turk_yt_2011}.
The approach usually followed is to project data on a fixed resolution grid, which is then used to compute gradients using a finite-difference scheme.
Even though this approach yields sensible results when the fixed grid resolution matches the AMR resolution, it yields null values when the AMR resolution is coarser and smoothes out the gradients when the AMR resolution is finer.
Therefore, we will deal here directly with the physical quantities at the scale of the AMR grid.

Using a tree search algorithm, we have developed a post-processing tool that is able to compute finite-difference gradients directly on the AMR grid.
The binary search algorithm ensures that any given location is found in at most $N$ steps, where $N$ is the number of AMR levels in the simulation (typically between 10 and 20).
To do so, we have augmented the yt code \citep{turk_yt_2011} to enable the computation of gradients for octree AMR datasets (The yt Project, in prep.).
The algorithm works as follows. (a) Loop over all octs in the tree.
(b) Compute the positions of the $4^3=64$ virtual cells centred on the oct and extending in $\pm 2\Delta x$ in three directions.
(c) Get the value of interest at the centre of each virtual cell from the AMR grid. If the virtual cell exists on the grid or is contained in a coarser cell, the value on the grid is directly used. If the virtual cell contains leaf cells, the mean of these cells is used.
(d) Compute the gradient of the quantity using a centred finite-difference scheme on the $4^3$ grid.
(e) Store the value of the gradient in the central $2^3$ cells.

\begin{figure*}
    \centering
    \includegraphics[width=\textwidth]{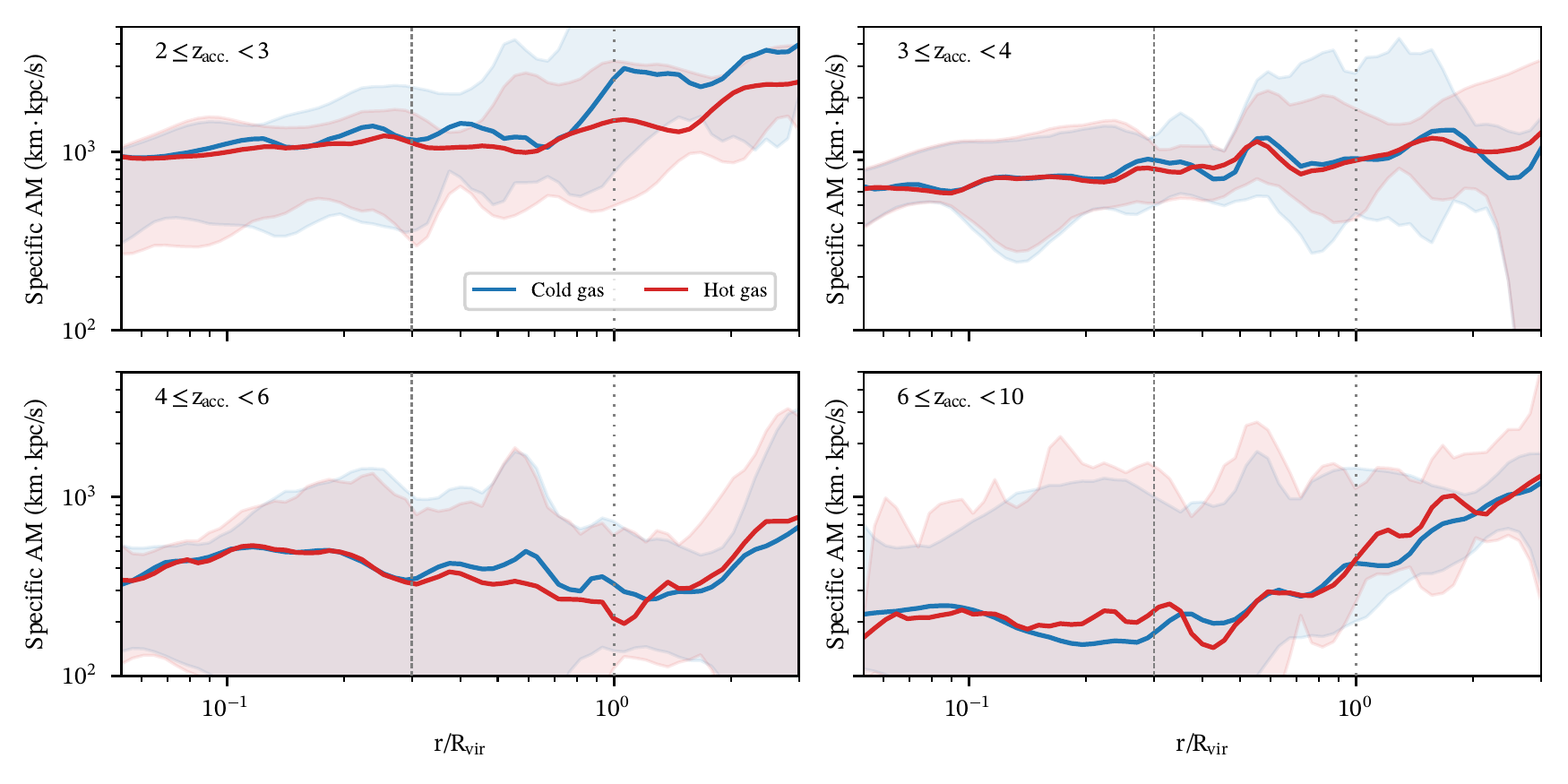}
    \caption{
    Radial profile of the \sAM in spherical shells in the cold phase (blue lines) and hot phase (red lines) for different redshift of accretion (each panel, as labelled).
    Error bars (shaded regions) are estimated from the standard deviation of the six galaxies.
    We show the virial radius (dotted gray line) and the edge of the CGM (dashed gray line).
    At $2 \leq z < 3$, the cold gas enters the virial radius with a larger specific angular momentum than in the hot phase.
    Once it reaches the CGM ($R_\mathrm{vir}/3$), the typical specific angular momentum is at all time comparable between the cold and the hot phase.
    }
    \label{fig:sAM-vs-r}
\end{figure*}

\begin{figure*}
    \centering
    \includegraphics[width=\textwidth]{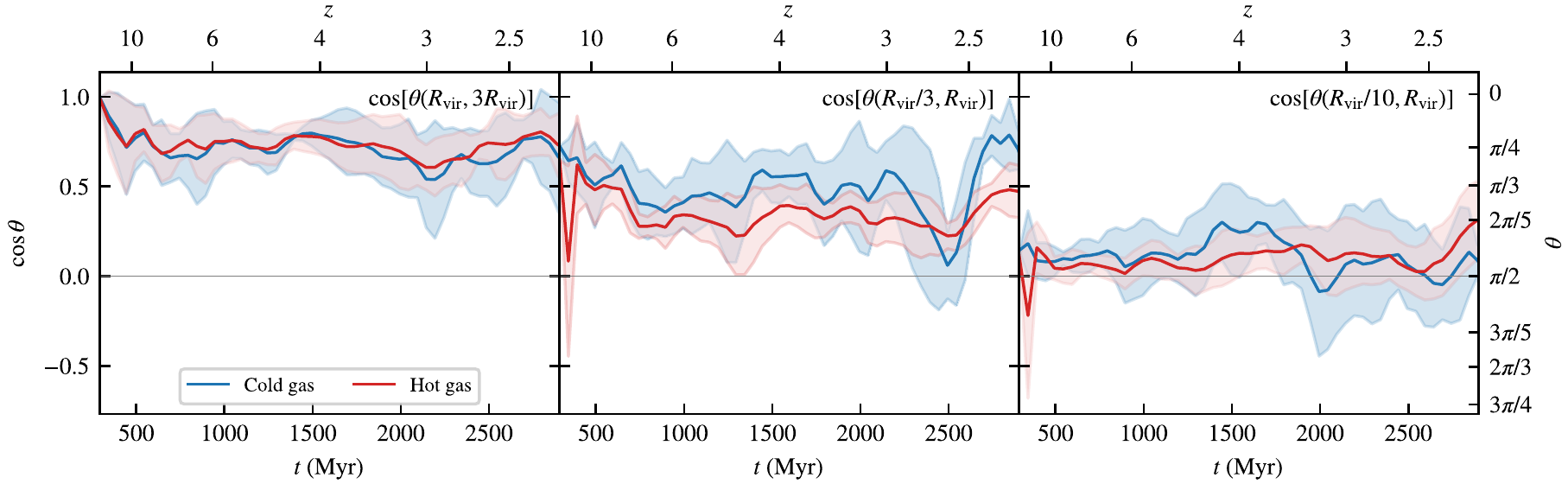}
    \caption{
    Thick line: median relative orientation of the \sAM of the cold gas at $\Rvir$ compared to its value at $3\Rvir$ (\emph{left}), at the edge of the CGM, $\Rvir/3$ (\emph{middle}) and at the disk's edge, $\Rvir/10$ (\emph{right}) in the cold phase (blue lines) and hot phase (red lines).
    Error bars (shaded regions) are estimated from the standard deviation of the six galaxies.
    The orientation of the \sAM is mostly conserved in the cold and hot phase between $3 R_\mathrm{vir}$ and $R_\mathrm{vir}$.
    Between the virial radius, $R_\mathrm{vir}$, and the CGM, $R_\mathrm{vir}/3$, the orientation of the cold phase is better preserved than in the hot phase.
    This orientation is mostly lost between the CGM and the disk, $R_\mathrm{vir}/10$.
    }
    \label{fig:sAM-relative-orientation}
\end{figure*}
\subsection{Cold gas selection}
\label{sec:numerical.cold-gas-selection}

The ratio of the total accreted mass with a maximum temperature below a given threshold $T_\mathrm{max}$ to the total gas mass --~the cold fraction~-- is a widely reported quantity in the study of the cosmological gas accretion, dating back to \cite{keres_how_2005}.
In this paper, we use a constant temperature cut $T \lesssim T_\mathrm{max}=\SI{2.5e5}{K}$ \citep[see \cref{appendix:temperature_threshold} for a discussion on the effect of the threshold, see also][]{nelson_moving_2013}.
We define the `cold gas' or `cold-accreted gas' at redshift $z_0$ as the particles that match all three criteria:
\begin{enumerate}
    \item the baryons have been accreted by $z_0$, \emph{i.e.} they are within the inner region of the halos $r<0.3 R_\mathrm{vir}$ at $z=z_0$;
    \item prior to  accretion, the baryons were in the gas phase and never heated above the threshold temperature $T_\mathrm{max}$ from $1.5 R_\mathrm{vir}$ to $0.3 R_\mathrm{vir}$,
    \item the baryons were never accreted on a satellite galaxy prior to their accretion in the main galaxy.
    In practice, this is done by excluding any particle found at any time at less than a third of the virial radius of any halo other than the main one.
\end{enumerate}
Conversely, particles that heated up at least once above the temperature threshold during their accretion will be referred to as `hot gas' or `hot-accreted gas'.
{We stress that in the selection steps (i) and (iii), we include baryons in all phases, including in the stellar phase. Thanks to the Monte-Carlo tracer particle scheme we use, these baryons can be accurately traced back in time, including when moving from one phase to the other \citep[see][for more details]{cadiou_AccurateTracerParticles_2019}.
We also emphasise that the temperature cut in step (ii) relies on the complete Lagrangian thermal history of the gas as provided by the tracer particles, rather than its instantaneous temperature.}
We have checked that the results obtained in this paper are robust to the exact temperature threshold used, as shown for the cold gas fraction in \cref{appendix:temperature_threshold}. We also checked that the results presented in the following sections are robust with respect to the particular choice of temperature threshold.

This cold gas is accreted along a clear filamentary structure, which is illustrated for galaxy~A in \cref{fig:3drendering}.
Following \cite{keres_how_2005}, we compute the ratio of the cold gas accreted to the total accreted gas as a function of time, which we show on \cref{fig:cold_gas_fraction}.
We recover the expected behaviour that the fraction of cold gas drops below redshift $z=3-4$ when the halo grows past $M_{\rm vir}\simeq 10^{11}\,\rm M_\odot$,
as expected due to the shock-heating of cosmic cold flows \citep{dekel_galaxy_2006}.

\section{Results: relative torquing}
\label{sec:numerical.results}

\subsection{The angular momentum magnitude}
\label{sec:AM_magnitude}
Before turn-around, gas acquires \AM \emph{via} torque from the cosmic web as captured by {TTT} \citep{hoyle_cosmological_1949,peebles_origin_1969,white_AngularMomentumGrowth_1984,catelan_EvolutionAngularMomentum_1996}.
The subsequent evolution of the cold-accreted and hot-accreted gas may however differ.
In order to study how the \sAM evolves, one can study the Lagrangian evolution of the \sAM of all gas accreted at the same time as a function of its radius, as shown on \cref{fig:sAM-vs-r}, and depending on its accretion mode.
The figure presents the Lagrangian evolution of the \sAM as a function of radius for the cold (blue lines) and hot gas (red lines).
In all halos, the \sAM of the gas decreases by a factor $\sim 4$ between $2\Rvir$ and $\Rvir /3$ in both accretion modes (cold and hot). We however note that at late times $2<z < 3$, the \sAM of the cold-accreted gas prior to its accretion is about twice larger than the hot-accreted gas, yet it ends up with a comparable one in the disk.

\subsection{The angular momentum orientation}
\label{sec:AM_orientation}
So far, we have only described the evolution of the magnitude of the \sAM of the gas.
In practice, the evolution of the orientation of the \sAM evolves slightly differently.
In order to quantify the evolution of the \sAM orientation, we define for a parcel of gas the angle between its \sAM at radius $R_1$ and its \sAM at radius $R_2$ as
\begin{equation}
    \cos \theta = \frac{\vec{l}(R_1) \vdot \vec{l}(R_2)}{\norm{\vec{l}(R_1)}\norm{\vec{l}(R_2)}}.
    \label{eq:costheta}
\end{equation}
We emphasize here that this angle is measured individually for each tracer particle using its Lagrangian history, so that $\vec{l}(R_1)$ may be measured at a different time than $\vec{l}(R_2)$.
Values close to one are found if the orientation is preserved, whereas random reorientations would yield values close to zeros.
We measure $\cos\theta$ as a function of the time of crossing $\Rvir$ for all particles in the cold- and hot-accreted gas in our six galaxies. We then compute the mean $\langle\cos\theta\rangle$  in the respective phase of each  galaxy, and also show in \cref{fig:sAM-relative-orientation} its median value from the sample of six galaxies.
We show $\cos\theta$ measured between $\Rvir$ and $3\Rvir$ (left panel), between $\Rvir$ and the outer edge of the CGM ($\Rvir/3$, centre panel) and between $\Rvir$ and the disk ($\Rvir/10$, right panel).
In all three panels, the reported time is always the time at which the particles crossed the virial radius.
We see that most of the realignment happens in the CGM (right panel), where $\cos\theta$ drops to values close to -- yet weakly greater than~-- 0.
Before entering the halo, the evolution of the orientation of the \sAM of the hot gas is similar to that of the cold gas: the orientation is conserved from $3\Rvir$ to $\Rvir$ but it becomes significantly less aligned than the cold gas between $\Rvir$ and $\Rvir/3$, where the misalignment is typically of the order of $2\pi/5$ ($\sim \SI{70}{\degree}$), compared to $\pi/3$ for the cold gas.
We do not report any significant evolution of the \sAM orientation with redshift.

\begin{figure}
    \centering
    \includegraphics[width=\columnwidth]{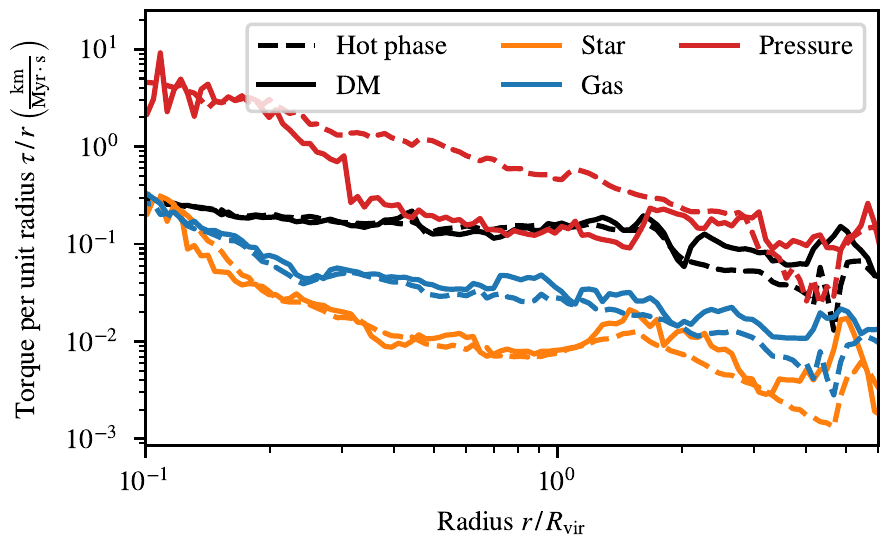}
    \caption{
    Mean radial profiles of the magnitude of the torque per unit radius of the DM (black), stars (orange), gas (blue) and pressure (red) in the cold phase (solid lines) and in the hot phase (dashed lines) at $z=3$.
    \emph{Local} torques are dominated up to $\approx 2R_\mathrm{vir}$ by pressure torques.
    }
    \label{fig:radial_profile_acceleration}
\end{figure}

\begin{figure*}
    \centering
    \includegraphics[width=\textwidth]{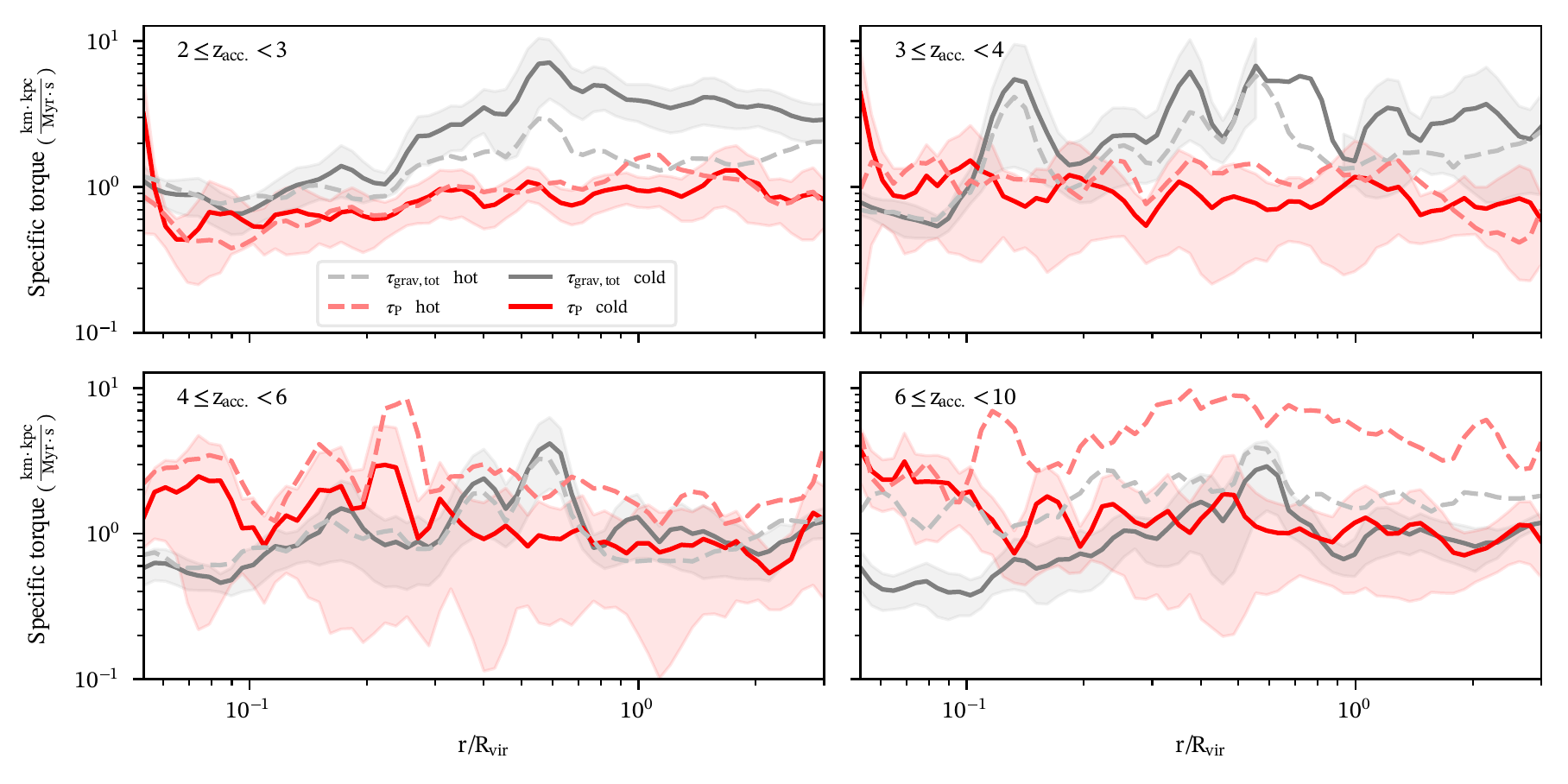}
    \caption{
        Mean angular-averaged profiles of specific gravitational torques (black) and specific pressure torques (red) integrated in spherical shells in different redshift bins.
        Compared to \cref{fig:radial_profile_acceleration}, we compute here the magnitude of the mean of the torques, $\norm{\sum\vec{\tau}}$, rather than the mean of the magnitudes, $\sum\norm{\vec{\tau}}$ ; this yields the net contribution of each torques applied to each gas phase.
        The cold accreted gas (solid line) and the hot accreted gas (dashed lines) are computed for all six halos, the lines indicate the mean value form the sample of six halos and the shaded region show the standard deviation.
        At $z\leq 4$, gravitational torques weakly dominate globally in the cold phase while at $z>6$, pressure torques dominate in the hot phase.
}%
    \label{fig:pressure-torques-cold-hot}
\end{figure*}

\begin{figure*}
    \centering
    \includegraphics[width=\textwidth]{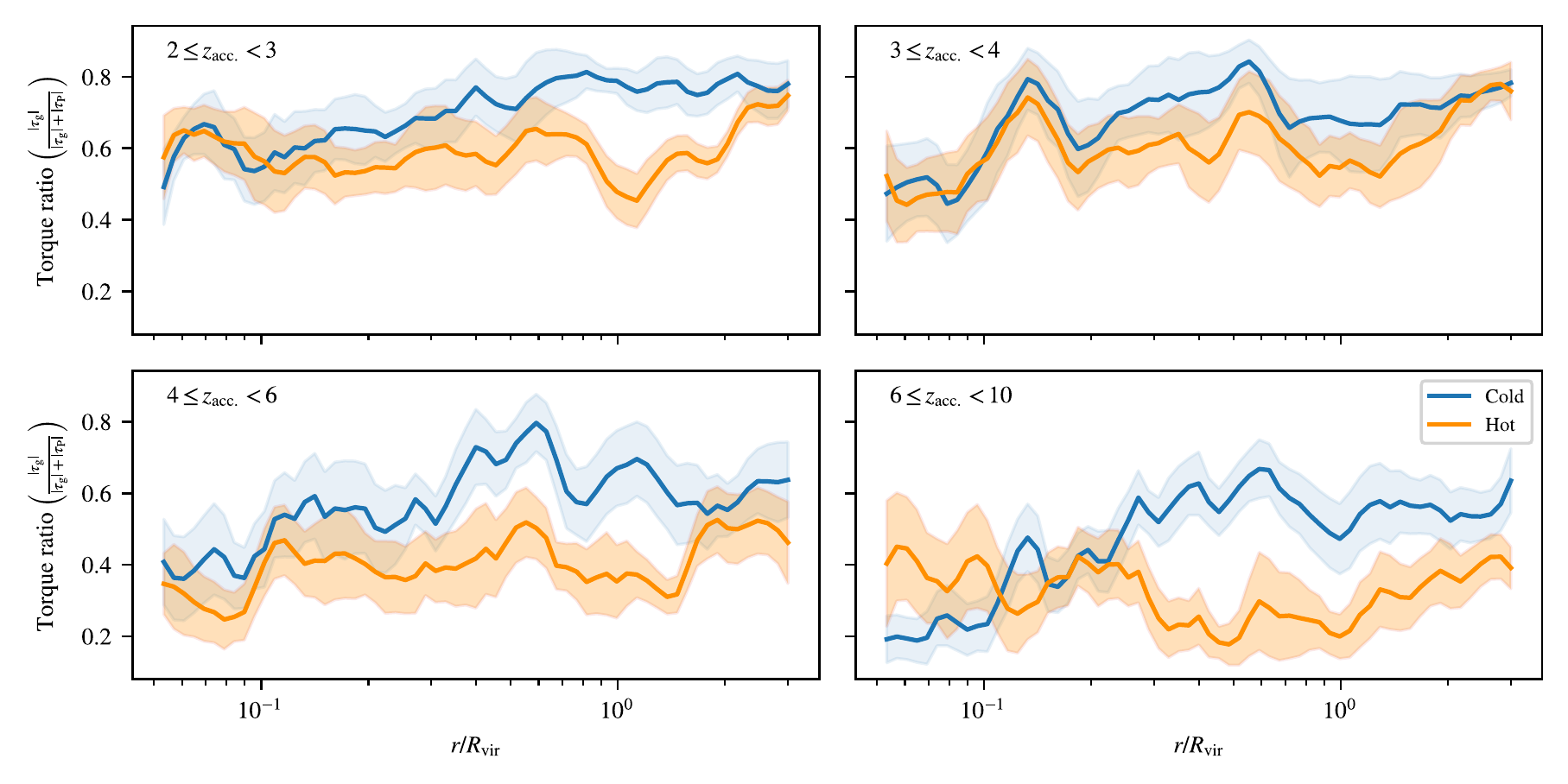}
    \caption{
       Ratio of the magnitude of the angular-averaged gravitational torques to the total torques in the cold phase (blue) and hot phase (orange). Where gravitational torques dominate, this ratio is close to 1. Where pressure torque dominate, it is close to 0.
       Importantly, gravitational torques prevail  ($>\SI{50}{\percent}$) throughout the halo in the cold phase at all redshifts and in the hot phase at $z\leq 4$.
    }%
    \label{fig:pressure-torques-ratio}
\end{figure*}

\begin{figure*}
    \centering
    \resizebox{0.75\textwidth}{!}{%
        \includegraphics[align=c,height=3cm,trim={0 0 1.132cm 0},clip]{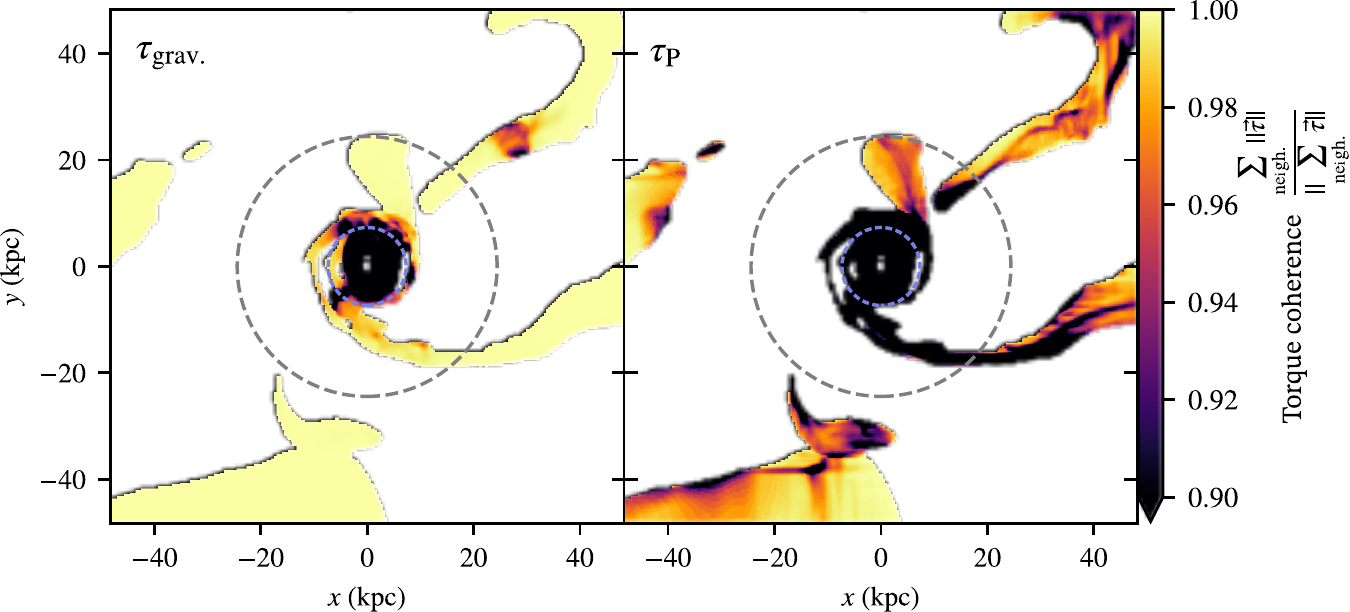}
     }
     \rotatebox[origin=c]{90}{
        \hspace{1em} Torque coherence $\displaystyle\dfrac{\norm{\sum_\mathrm{neigh} \vec{\tau}}}{{\sum_\mathrm{neigh}\norm{\vec{\tau}}}}$}
    \caption{
    Projected map of the coherence of the gravitational torque and pressure torques (left and right panel respectively) around galaxy~A at $z=2$ in the cold gas phase.
    The effective smoothing scale is \SI{1}{kpc} (pixel size).
    The virial radius is shown as a dashed line, a third of the virial radius is shown as a blue dashed line.
    Gravitational torques are coherent in the cold flows down to the edge of the CGM, while pressure torques very much less so.
    A qualitatively similar picture emerges for similar maps in the hot phase.
    }
    \label{fig:coherence_of_torques}
\end{figure*}

\begin{figure}
    \centering
        \hspace{.5em}
        \rotatebox[origin=c]{90}{
      \hspace{1em} Torque coherence $\displaystyle\dfrac{\norm{\sum_\mathrm{neigh} \vec{\tau}}}{{\sum_\mathrm{neigh}\norm{\vec{\tau}}}}$}
        \hspace{.5em}
        \includegraphics[align=c,height=4.75cm,trim={1.08cm 0 0 0},clip]{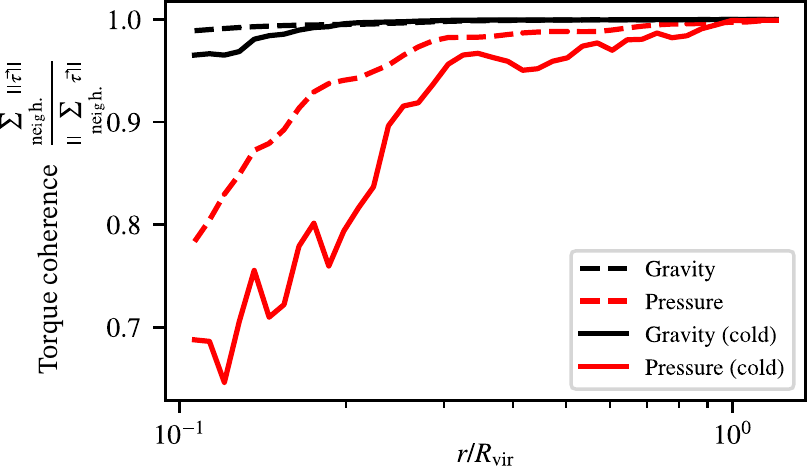}
    \caption{
    The radial profiles of the coherence of torques in the cold (solid) and hot (dashed) phases for the pressure torques (red) and gravitational torques (black).
    The coherence is defined as the ratio of the norm of the sum $\norm{\sum_\mathrm{neigh} \vec{\tau}}$ to the sum of the norms $\sum_\mathrm{neigh} \norm{\vec{\tau}}$
     in a cell and its 26 neighbours.
    Gravitational torques are coherent over large scales, so that the sum of torques exerted on neighbouring gas cells adds up.
    On the contrary, pressure torques fluctuate on the scale of individual gas cells, so that their sum partially cancels out.
    Pressure torques have particularly large fluctuations in the cold phase at small radii.
}
    \label{fig:coherence_of_torques2}
\end{figure}

\subsection{Dominant torques in the cold and hot phase}
\label{sec:dominant-torques}

We have presented in \cref{sec:AM_magnitude,sec:AM_orientation} that the cold gas retains its orientation and magnitude down to the CGM, while the hot gas starts realigning at a larger radius.
Let us now study which torques are responsible for the spin-down and realignment of the gas.
We split the total torques between pressure torques and gravitational torques, as was done in \cref{eq:sAM-evolution}.
The gravitational torques are further split between their DM, gas and stellar components following \cref{sec:numerical.methods.grav_torques} in order to assess the role of the various components.
We then sum the specific torque exerted by each of the component on the gas in concentric shells and mass-average (using the gas tracer particle mass) it to obtain `angular-averaged profiles'.

We show on \cref{fig:radial_profile_acceleration} angular-averaged profiles of the magnitude of each of the torque components, $\langle\norm{\vec{\tau}}\rangle$, averaged over all six galaxies, for the gas accreted at $z=3$ or earlier\footnote{Because our criterion to define cold/hot accretion depends on the entire thermal history during accretion, the analysis of the gas at two virial radii need to be done at least two free-fall times before the end of the simulation ($z=2$) for it to have time to be accreted by the central galaxy.}.
The profiles are computed independently for the cold and the hot phase.
Since we average the magnitude of torques\footnote{As opposed to taking the magnitude of the averaged torque, as  done later.}, this allows us to quantify which torques dominate  \emph{locally} the evolution of the angular momentum.
First, we observe that the radial profiles of the gravitational torques are comparable in both phases of the gas.
This is expected since gravity is a non-local force and is thus independent of the local thermodynamical conditions of the gas.
On the contrary, pressure torques, which depend on the local thermal structure of the gas, display significant differences between the cold and the hot phase.
Between $2 \Rvir$ and $\Rvir/3$, pressure torques in the hot phase are on average one order of magnitude larger than in the cold phase.
We find no evidence of a difference in the torques (both gravitational and pressure) in the two phases prior to entering the halo and once inside the CGM.

Gravitational torques are driven by DM all the way down to the galactic disk at $\Rvir/10$, where the contribution from the stellar disk and the gaseous disk become comparable.
In the cold phase, we find that pressure torques have the same magnitude as the DM gravitational torques in the circum-galactic medium  and outside the halo, while they dominate by an order of magnitude in the hot phase.
Let us recall that the origin of pressure torques are local by nature, while gravitational torques are sourced by the full matter distribution within the halo scale.
As a consequence, we expect that gravitational torques are more coherent, \ie their effect adds up, over large scales and over time.
On the contrary, we expect pressure torques to have small-wavelengths short-timescales variations, so that their net contribution to torquing the gas may partially cancel out when averaged over large scales, hence over time when following the flow.
To quantify this, we compute in \cref{fig:pressure-torques-cold-hot} the magnitude of the angular-averaged torques, $\norm{\langle\vec{\tau}\rangle}$, in the cold and hot phases, as a function of accretion time.
We note that the gravitational torques acting on the cold/hot accreted gas have the following  hierarchy: DM gravitational torques dominate down to $\Rvir/10$ where they become comparable to torques from the stellar disk. Interestingly, torques from gas are not dominant anywhere from $2\Rvir$ to $\Rvir/10$.
For the sake of clarity, we only show the total gravitational torques, $\vec{\tau} = \vec{\tau}_\mathrm{DM} + \vec{\tau}_\star + \vec{\tau}_\mathrm{gas}$ instead of each individual component.
We find that the \emph{net} gravitational torques dominate over the \emph{net} pressure torques in the cold phase (solid lines) at $z \leq 4$, and slightly less so in the hot phase (dashed lines), in the CGM.
At high-redshift $z>6$, this hierarchy reverses in the hot phase, which becomes dominated by pressure torques.
Let us however note that at these redshifts, only a marginal fraction of the matter is accreted hot, as was shown on \cref{fig:cold_gas_fraction}, so that the hot-accreted gas may be a highly-biased \mbox{(sub-)sample}.
We also checked that the hierarchy of gravitational torques is the same as in \cref{fig:radial_profile_acceleration}: DM gravitational torques dominate gravitational torques outside the galactic disk, and baryons dominate inside of it.

With the same redshift cuts as in \cref{fig:pressure-torques-cold-hot}, we finally show in \cref{fig:pressure-torques-ratio} the ratio of the net gravitational torques to the sum of the net torques in the cold and hot phases.
This allows us to make more quantitative statements about the net torques in the cold and hot phases.
Below $z<4$, we report that gravitational torques are responsible for the majority of the torquing in the CGM and down to the edge of the disk in the cold and hot phase.
In the cold phase, gravitational torques dominate at all times in the CGM, while in the hot phase, pressure torques dominate until $z>4$; at $z<4$, the dynamical evolution becomes marginally dominated by gravitational torques.

\subsection{Spatial and temporal coherence of  torques}
\label{sec:coherent-torques}

We have seen that, locally, pressure torques dominate over gravitational torques in the hot phase, and have comparable magnitudes in the cold phase.
However, once averaged over all angles, the latter dominate over the former at late times.
Let us now investigate the origin of this reversal by estimating the coherence of the torques.
In principle, the spatial coherence would be best captured by computing the power spectrum of each of the torques, but the intricate geometry of the cold and hot phases render this approach complex.
Instead, we rely on a simple estimator of the spatial coherence of the torques on small scales. To do so, we estimate the torques on a regular Cartesian grid and compare the magnitude of the sum of the torques in a cell and its $3^3-1=26$ neighbours to the sum of the magnitudes, \ie
\begin{equation}
    r_\alpha \equiv \frac{\sum_{i=1}^{27} \norm{\vec{\tau}_{\alpha,i} } }{\norm{\sum_{i=1}^{27}\vec{\tau}_{\alpha,i}}},
\end{equation}
where $\alpha$ indicates which component we are interested in (gravitational, pressure) and $i$ runs over the indices of the central cell and its 26 neighbours.
The grid we project onto has cells of size \SI{500}{pc} and we thus measure fluctuations in the torques on $\sim \SI{1}{kpc}$ scales.
Ratios close to one indicate regions where the torques adds up, while ratios close to zero are found where torques lack spatial coherence and cancel out.

We show in \cref{fig:coherence_of_torques} %
mass-weighted projected maps of the coherence ratio, $r_\alpha$, in the cold phase.
Gravitational torques are coherent all the way to the CGM, while pressure torques much less so.
This is quantified in \cref{fig:coherence_of_torques2}, %
in which we show radial profiles of the ratio in both phases for the pressure and gravitational torques.
This illustrates that pressure torques have little spatial coherence, as expected, so that different locations of the cold flows may be either spun-up or spun-down.
On the contrary, large patches of the cold flows undergo coherent gravitational torques that can add up.
In the disk, all torque sources lose their long-range spatial coherence and appear noisy.
\section{Discussion}
\label{sec:numerical.discussion}

\subsection{Comparison to Danovich et al. (2015)}
\label{sec:numerical.results.sAM-vs-per-volume}

This work differs from~\cite{danovich_four_2015} as follows (a) we make use of simulation with a similar maximum resolution of $30\ \rm pc$ but with a different code and different sub-grid models, (b) we rely on tracer particles which trace the history of gas elements, (c) we compute the gravitational torques from each component, (d) we compute the pressure torques directly in the AMR grid, and (e) we refer here to the AM per unit mass.

First, taken at face value, our results of \cref{fig:sAM-vs-r} appear to be in conflict with their Figure 1, that showed that the cold gas has a spin parameter $2-3$ larger than the hot gas at the virial radius.
Indeed, we find that both the cold and hot gas have comparable spin parameters upon entry in the virial radius, except for the gas accreted at $z\lessapprox 3$, where we see hints that the cold gas has a specific angular momentum larger by a factor $\approx 2$.
The main difference is that we represent here the \emph{past} angular momentum of the gas accreted by a given time, as a function of its past radial location.
If we are to compare our results to those of \cite{danovich_four_2015}, we thus need to take into account the fact that the gas will take an additional $\approx 500\ \mathrm{Myr}$ to go from the virial radius to the outer edge of the CGM, i.e.\ the value of the angular momentum at $\Rvir$ in our top-left panel is for gas at roughly $3.8 \gtrapprox z \gtrapprox 2.4$, so it should be compared to the left and central panels of~\cite{danovich_four_2015}, Figure 1, where a similar ratio ($\approx 2$) is observed.

Second, \cref{eq:sAM-evolution} differs from their equation~9, which includes a term proportional to the velocity divergence. Our analysis relies on the analysis of the \sAM instead of the \AM per unit volume.
The fundamental difference is that the rate of change of \AM per unit volume includes a dependence to the cell volume \emph{via} the velocity divergence.
The latter is itself highly sensitive to the compression or expansion of the gas, which are ubiquitous in highly-compressible astrophysical flows.
Contrary to what \cite{danovich_four_2015} reported, we find that the velocity divergence term dominates over the gravitational and pressure terms.
Indeed, inflowing gas typically moves at \SI{100}{km.s^{-1}} with typical variation scales of a few \si{kpc}.
An estimate of the magnitude of the velocity term yields $\norm{l\div{\vec{v}}} / l \approx  \SI{100}{km.s^{-1}} / \SI{1}{kpc} \approx \SI{100}{Gyr^{-1}}$, with larger values found in shocked and highly compressed regions.
This has to be compared to torques we measured, $\tau / l \approx (\SI{2}{km.s^{-1}.kpc.Myr^{-1}})  / (\SI{1000}{km.s^{-1}.kpc}) \approx \SI{2}{Gyr^{-1}} \ll \norm{l\div{\vec{v}}} / l$.
In this paper, we have avoided this issue by studying the \sAM, whose evolution is described by \cref{eq:sAM-evolution}, in which no velocity divergence appears.
Since our conclusions are similar to those of~\cite{danovich_four_2015}, it appears that the effect of velocity divergence on angular momentum evolution does not play a significant role, even when considering the angular momentum per unit volume.
Furthermore, the \sAM of tracer particles -- which sample that of the gas -- can be readily computed from their trajectories and velocities, and their \AM is simply obtained by multiplying by the fixed tracer particle mass.

Overall, we find that the cold gas accreted by $z<3$ has more angular momentum than the hot gas prior to accretion (phase I of \cite{danovich_four_2015}), yet at $z>3$, we do not find a significant difference. The angular momentum is then transported with a roughly constant magnitude (phase II). While we do not find a significant drop of the magnitude of the angular momentum of the cold and hot gas once in the CGM at $\Rvir/3$ (phase III), we however notice that it reorients itself in this region.
We suggest future work be focused on resolving correctly the CGM phase, with a special emphasis on understanding how it interfaces between cosmological accretion, CGM and outflows from feedback processes.

\subsection{Impact of gravitational interactions and role of the CGM}

At large radii, the evolution of the \AM follows the tides imposed by the cosmic web, as explained by {TTT} \citep[\eg][]{codis_connecting_2012,codis_spin_2015}.
The gas then flows on the forming galaxy \emph{via} two different channels: the hot and cold accretion modes, in particular for massive enough galaxies at $z \gtrsim 2$ \citep{birnboim_virial_2003,keres_how_2005,dekel_galaxy_2006,ocvirk_bimodal_2008,nelson_moving_2013}.
The predominance of one or the other channel can be used to understand the formation of disky galaxies and the internal evolution of the galaxy.
Indeed, in cold flows, the gas is able to penetrate deep in the halo and can feed the galaxy with fresh gas: we confirm that it does so with a steady \AM orientation down to the CGM ($\Rvir/3$), in agreement with previous findings \citep{pichon_rigging_2011,stewart_angular_2013}.

In numerical simulations, it has been observed that cold gas has a higher \AM at larger radii, as measured by their spin parameter \citep{kimm_angular_2011,tillson_angular_2015,danovich_four_2015} which is up to one order of magnitude larger than that of the DM.
However, in the CGM and the disk, the spin parameter of the gas is found to be only three times larger than that of the DM at the same location, and the nature of the torques leading to this spin-down is still debated today.
\cite{danovich_four_2015} argued that the dominant torques, in halos of mass $\sim10^{12}\, \rm \Msun$ at $1.5<z<4$, are gravitational torques regardless of the distance to the galaxy.
Focusing on a massive halo of $3\times 10^{10}\,\rm \Msun$ at $z\geq 6$, \cite{prieto_how_2017} found that, for those objects, the dominant torques were from pressure torques.
In this work, we find that pressure torques are locally dominant in the hot-accreted gas and are comparable to the DM gravitational torques in the outer halo.
While pressure forces can act locally as the dominant forces, we found that their \emph{net} contribution is subdominant in the cold phase until $z=2$, and in the hot phase until $z\approx 4$.
We found that this reflects the lack of large-scale coherence of pressure torques, which cancel out when averaged over large scales.
Conversely, gravitational forces, which depend on the distribution of matter on larger scales, are able to coherently torque the infalling gas, resulting in most of the spin-down signal.

The net effect of the gravitational forces is to spin-down the accreted gas.
Indeed, non-spherical mass distributions are able to exchange angular momentum through gravitational interactions with the accreting gas.
One possible reason is the following: under the effect of gas infall, the DM halo and the disk becomes slightly elongated which in turn creates a tide that will torque the hot gas down.
In order to quantify the polarisation of the halo and disk induced by the gas in the halo and in the cold phase, one could decompose the gravitational potential created by these two components and verify it indeed creates such tidal features. This is however beyond the scope of the current paper.
Using the torque magnitudes of \cref{fig:radial_profile_acceleration}, the typical angular momentum of the gas upon its entry in the halo ($\sim \SI{e3}{km.s^{-1}.kpc}$) would be depleted in a time $t_{\tau,\mathrm{DM}}(R=\Rvir) \approx (\SI{e3}{km/s.kpc})/(\SI{2}{km.s^{1}.kpc.Myr^{-1}}) \approx \SI{500}{Myr}$, which is about one free-fall times $t_\mathrm{ff}=\SI{500}{Myr}$ at $z=2$.
In our simulations, we find that cold-accreted gas falls in about one free-fall time, while hot-accreted gas takes in average $1000\pm500\,\rm Myr$ to fall from $3\Rvir(z=2)$ to $\Rvir(z=2)/3$ where $\Rvir(z=2)$ is the final virial radius of the halo at $z=2$.
Since the hot gas lingers outside the CGM, torques, and in particular pressure torques, have more time to erase its initial direction set at cosmological scales. This induces a weaker alignment compared to the cold-accreted gas as we observed in \cref{fig:sAM-relative-orientation}.
Interestingly, even though the hot accreted gas lingers for longer, we note that its alignment with its pre-accretion direction is not entirely lost until $\Rvir/3$.

As reported in~\cite{rosdahl_extended_2012}, the trajectory of the cold gas is different and follows a mostly radial (with a non-null impact parameters) free-fall trajectory.
In our simulation, the cold gas typically takes $500\pm350\,\rm Myr$ to go from $3\Rvir$ to $\Rvir/3$, which is roughly equal to the typical torque-down timescale, so that cold-accreted gas has less time to be reoriented.
As the cold gas plunges into the halo, the influence of the disk increases up to the point where torques become dominated by stars.
Interestingly, the radius at which most of the \AM, orientation included, has been lost  coincides with the edge of the disk at $\Rvir / 10$.
This is in broad agreement with Figures 16 and 17 of~\cite{danovich_four_2015}, which found that the torques generated by the disk increase significantly at $\sim \Rvir / 10$, where they act to realign the angular momentum of infalling material to the plane of the disk spin.
This may be an indication that the disk is actually responsible for the realignment, and will require further work to understand the coupling between the accreted material and the stellar disk, the gaseous disk and the CGM.
In this scenario, the \AM of the freshly accreted material would align with the CGM spin and the galactic spin and, since this gas will later be mixed in the disk and eventually form star, it would provide an explanation for the observed alignment of the CGM's spin and the galactic spin measured in simulations, but not with the halo's spin as a whole~\citep{bailin2005,tenneti2015,velliscig2015,chisari_galaxy-halo_2017,2017arXiv171207818W,welker2018}.
This hypothesis could be addressed following a similar analysis as was done in this paper, but isolating further the gravitational torques from the CGM and from the galaxy alone: this will be the topic of future work.

\subsection{Effect of  cold flows stability on \AM transport}
\cite{cornuault_are_2018} proposed that cold flows do not survive within the halo.
They suggested that they instead fragment into (unresolved) clouds, while their internal pressure increases.
In the process, the kinetic energy of the gas funnelled in turbulence is dissipated as the result of radiating more thermal energy due to the more efficient mixing between the cold and hot phase, effectively loosing the shielding effect usually assumed for cold flows.
In this scenario, turbulence in cold flows could contribute to efficiently mix the angular-momentum rich cold gas to the hot gas.
This would likely result to a diffusion of the \AM of the cold gas into the hot medium and increase the relative importance of pressure torques to the problem of the \AM transport.

Using idealised simulations, \cite{mandelker_instability_2016}, \cite{padnos_instability_2018}, \cite{mandelker_instability_2019}, \cite{sparreetal19}, and~\cite{aung_KelvinHelmholtzInstabilitySelfgravitating_2019} showed that cold flows may be sensitive to the Kelvin-Helmholtz depending on redshift and the mass of the central object.
In particular, they showed that thin-enough filaments can be destroyed before reaching the galaxy.
In this last case, the cold gas would effectively lose its angular momentum to the hot halo before interacting with the galaxy.
However, the effect of gravitational torques would likely be similar, though they may act for a longer time before the gas plunges into the CGM.
Interestingly, these studies also suggested that cold flows may entrain the neighbouring hot gas as they fall in while slowing down the infall of the cold gas \citep{mandelker_InstabilitySupersonicCold_2020}, which may result in an efficient mixing of the \AM at the boundary of the cold flows.
\cite{berlok_impact_2019} suggested that the mixing may be decreased if one considers magnetised flows with field lines parallel to the flow, as a result of a magnetic tension working against the Kelvin-Helmholtz instability, while thermal conduction can also have an important effect on stabilising the mixing process~\citep{armillottaetal16,kooijetal21}.
Simulations in the cosmological with enhanced spatial resolution (with hyper-Lagragian refinement scheme) in the CGM have been realised but show contradictory results about the effect on the cold gas mass content of the CGM~\citep{vandevoort2019,peeples2019,suresh2019,hummels2019}, while the effect on the galaxy remains elusive.

\cite{dubois_blowing_2013} showed in a high redshift $z>6$ massive halo that the {AGN} activity is able to significantly decrease the cold gas mass reaching the CGM and modify the filamentary structure up to the outer regions.
\cite{nelson_impact_2015} showed that AGN feedback is able to significantly increase the infall time of the cold gas.
The overall effect is to increase the infalling time of the cold gas, which, if large enough, would allow torques to erase the initial direction of the cold-accreted gas, as is already the case for the hot accretion.

The exact role of the various feedback processes  on the transfers of angular momenta in the various CGM phases and onto the central galaxy will need to be elucidated: we defer this to future work.

\section{Conclusion}
\label{sec:numerical.conclusion}
\begin{figure*}
    \includegraphics[width=0.975\textwidth]{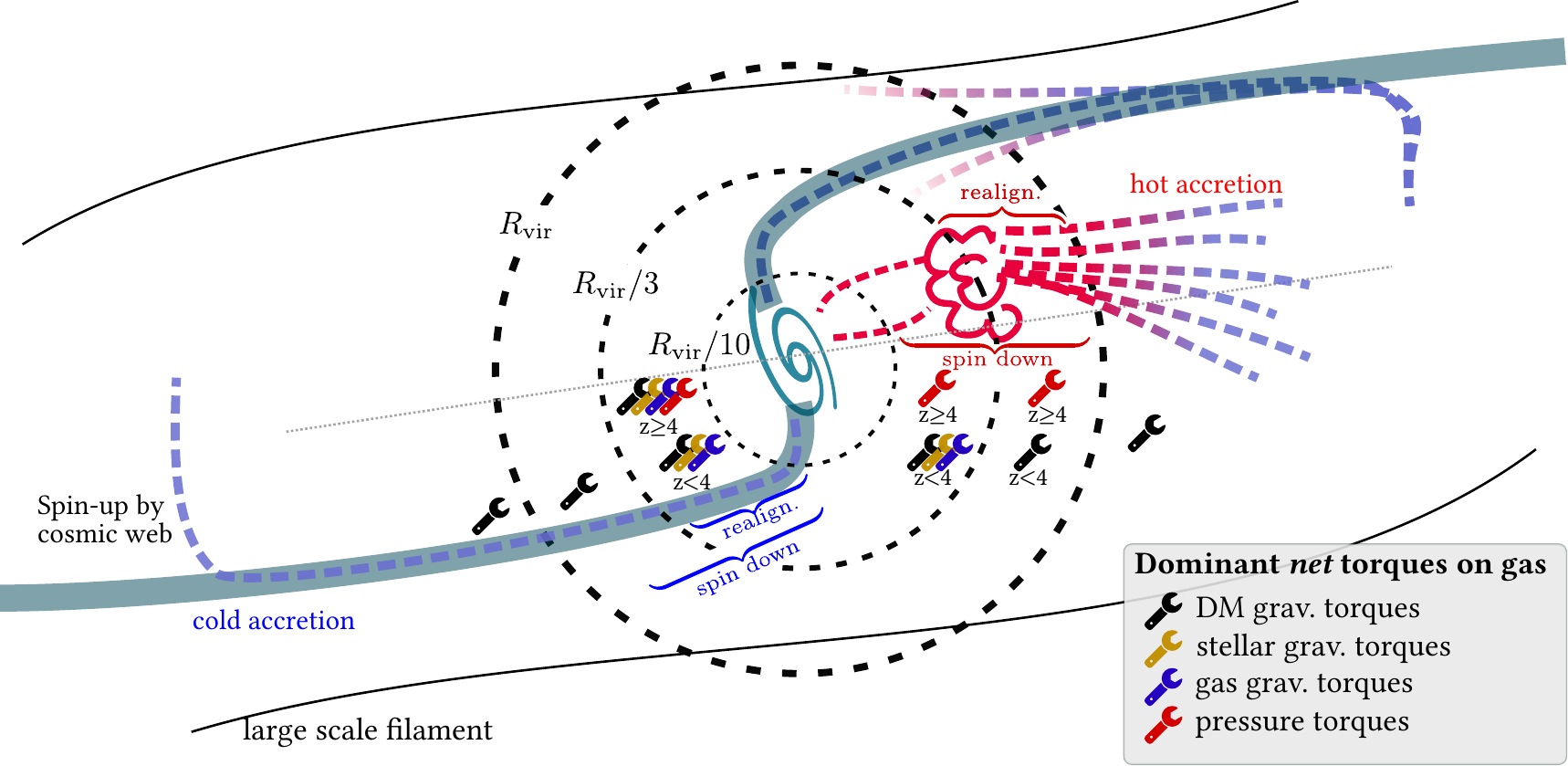}  %
    \caption{
        Summary sketch of the evolution of the \AM of the accreted gas and of the dominant torques applied on it.
        Outside the virial radius, DM torques dominate these torques.
        In the outer halo, DM torques dominate, except at high $z$ in the hot phase, where pressure torques dominate the dynamics.
        In the CGM, gravitational torques (DM, star and gas) dominate at $z<4$.
        At $z\geq 4$, pressure torque also contribute to the dynamics of the cold phase, while they dominate in the hot phase.
        These torques lead to the spin down of the accreted gas in the CGM, while the realignment of the AM happens at a closer radius for the cold-accreted gas.
        The CGM emerges as the interface between the large-scale cosmic web and the galactic disk where most of the realignment and AM redistribution happens.
    }
    \label{fig:galaxy-and-env}
\end{figure*}

Using a set of high-resolution zoom-in simulations including the physics for star formation, SN and AGN feedback from NewHorizon~\citep{duboisetal20}, we have studied the evolution of the \AM of gas accreted \emph{via} the cold and the hot mode around six group progenitors at $z\gtrsim 2$. We also presented new numerical methods to extract the contributions of the different forces and torques (gravitational and pressure torques).

We found that the mode of accretion has little impact on the evolution of the magnitude of the \sAM, but has a small but noticeable effect on its direction. Cold-accreted gas better retains its initial direction down to the CGM \citep[see also][their figure 8]{tillson_angular_2015}. This initial alignment is however mostly lost once the gas reaches the disk, regardless of its accretion mode.
We have provided a detailed analysis of the torques acting on the accreting material, revealing that pressure torques are \emph{locally} larger than gravitational torques in the hot-accreted phase, and are comparable in the cold-accreted phase.
Once averaged over all angles, we find that the \emph{net} DM gravitational torques dominate the global evolution of the \AM{} in cold gas, especially at later times. In the hot-accreted phase, however, the contribution of the gravitational torques is significantly reduced, and only dominates at $z<4$.

Our findings support models in which most of the \AM is able to flow down to the CGM where gravitational torques redistribute it to the DM and the disk component through gravitational interactions, effectively transporting \AM from the scales of the cosmic web into the CGM  \citep[][]{pichon_rigging_2011}.
At high-redshift, the emerging picture is that the spin of the gas, acquired at cosmological scales, is indeed transported mostly unimpaired in cold flows up to the CGM (see Fig.~\ref{fig:galaxy-and-env}).
Upon reaching the CGM, most of the spin's amplitude has been lost to other components, yet its orientation is still partly preserved (within \SI{60}{\degree}).
 A key qualitative difference of gas torquing specifically within the inner region of the CGM seem to be the growing impact of torques from the disc itself, which has its own local (partially de-correlated) orientation, reflecting the geometry of past accretion. As the contributions from such torques become dominant the flow finally flips one last time.
This highlights the crucial role played by the CGM interface between the cosmic web and the disk, which appears as the missing piece of the puzzle between the large-scale cosmological angular momentum acquisition, dictated by tidal torque theory, and the formation  of the embedded stellar disk, later shaped by secular evolution and feedback processes.

\section*{Acknowledgements}

This project has received funding from the European Union's Horizon 2020 research and innovation programme under grant agreement No. 818085 GM-Galaxies and the Agence nationale de la recherche grant Segal ANR-19-CE31-0017 (\href{http://secular-evolution.org}{http://secular-evolution.org}).
This work was granted access to the HPC resources of CINES under the allocation A0060406955 by Genci.
The analysis was carried out using
\textsc{colossus} \citep{diemer_COLOSSUSPythonToolkit_2018},
\textsc{jupyter} notebooks \citep{soton403913},
\textsc{matplotlib} \citep{hunter2007matplotlib},
\textsc{numpy} \citep{harris_ArrayProgrammingNumPy_2020},
\textsc{gnu parallel} \citep{tange_ole_2018_1146014},
\textsc{pynbody} \citep{pontzen_pynbody_2013},
\textsc{python} and
\textsc{yt} \citep{turk_yt_2011}.
This work has made use of the Infinity Cluster hosted by Institut d'Astrophysique de Paris. We thank Stephane Rouberol for running smoothly this cluster for us. We thank A.~Dekel, S.~White, D.~Pogosyan, J.~Devriendt and A.~Slyz for useful comments.
CC thanks the \textsc{yt} community and especially N.~Goldbaum and M.~Turk who contributed to make the analysis possible.

\section*{Data Availability}
The data underlying this article will be shared on reasonable request to the corresponding author.

\bibliographystyle{mnras}
\bibliography{biblio}

\appendix

\section{Impact of  temperature criterion}
\label{appendix:temperature_threshold}

\begin{figure*}
    \centering
    \includegraphics[width=\columnwidth]{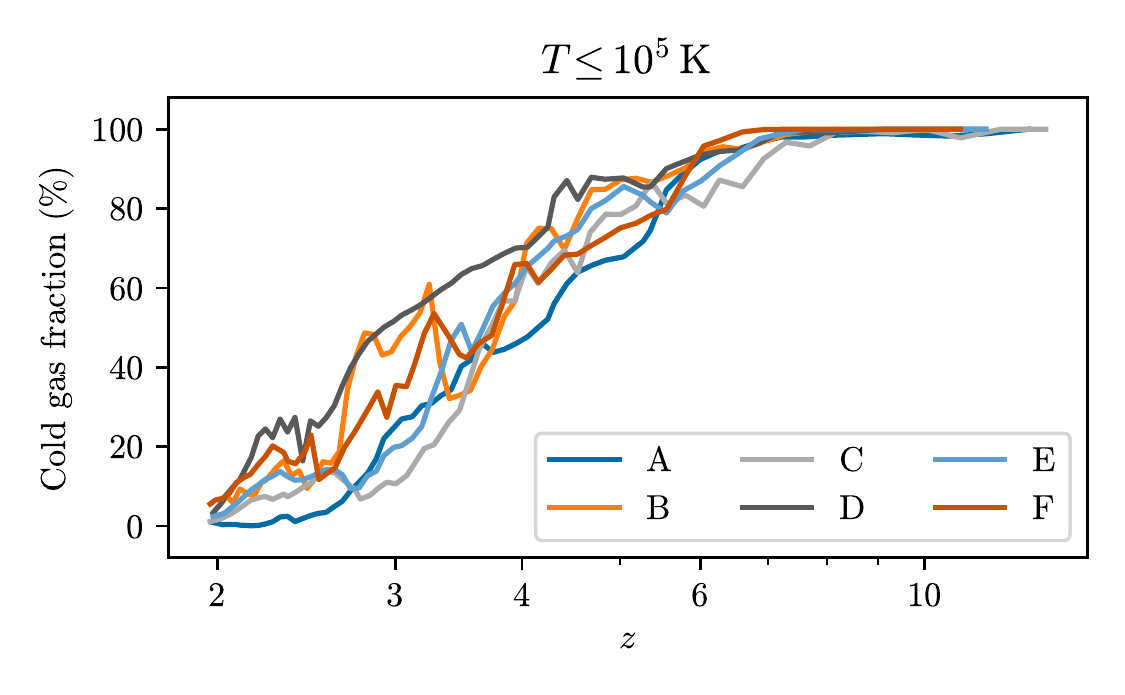}
    \includegraphics[width=\columnwidth]{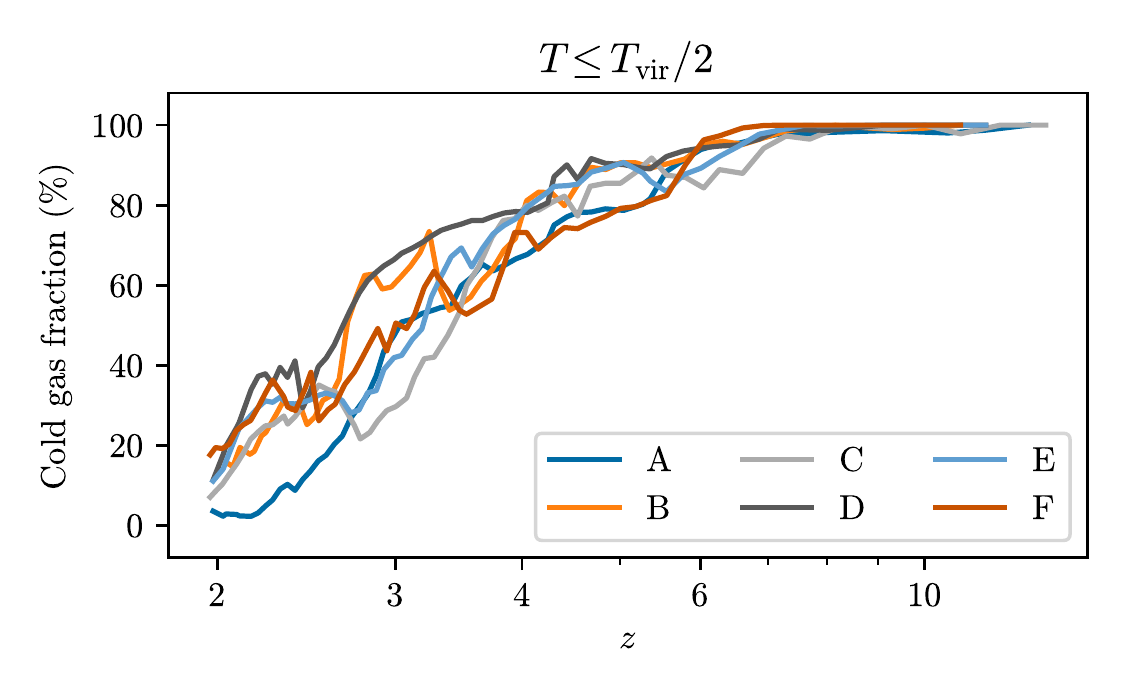}
    \includegraphics[width=\columnwidth]{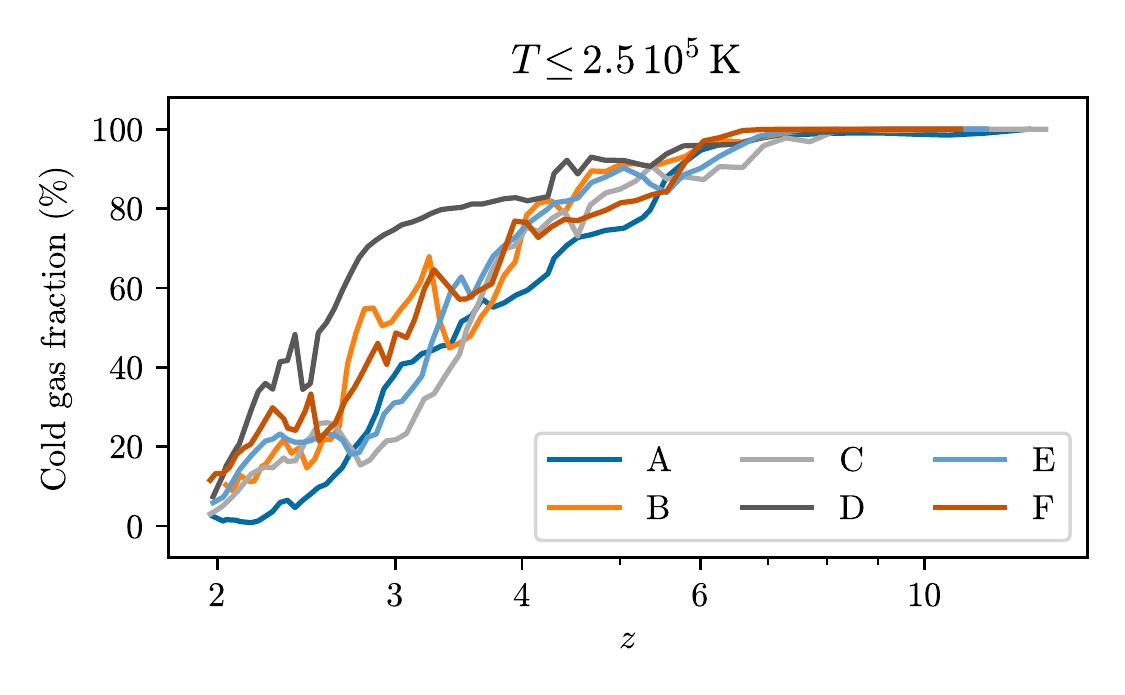}
    \includegraphics[width=\columnwidth]{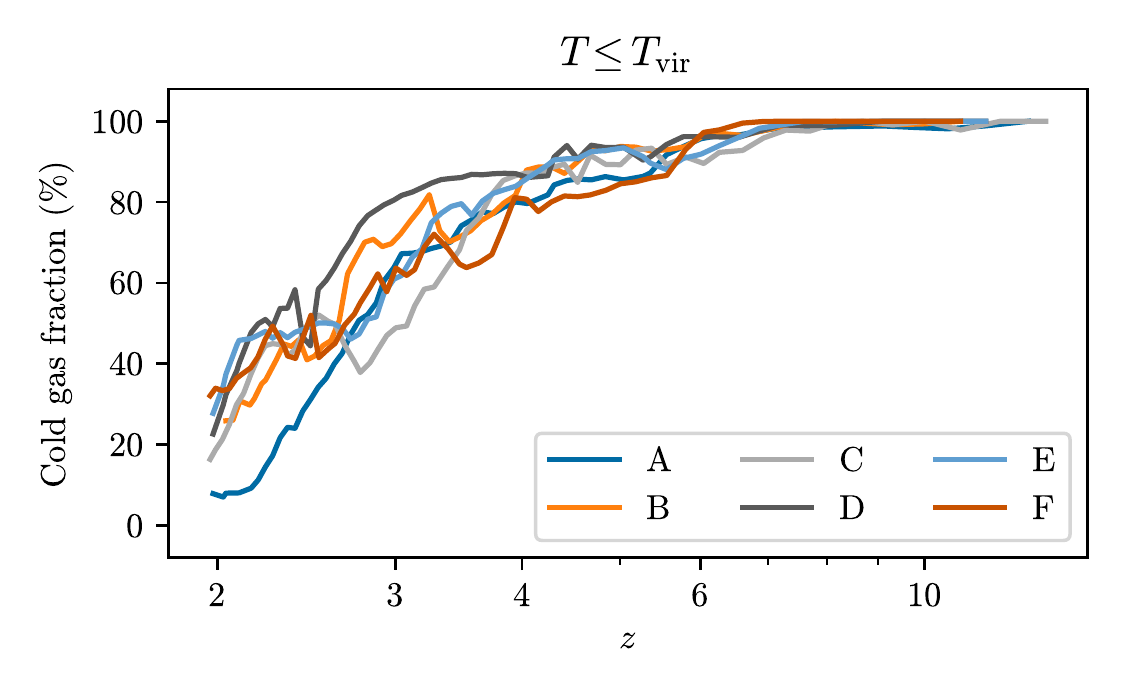}
    \includegraphics[width=\columnwidth]{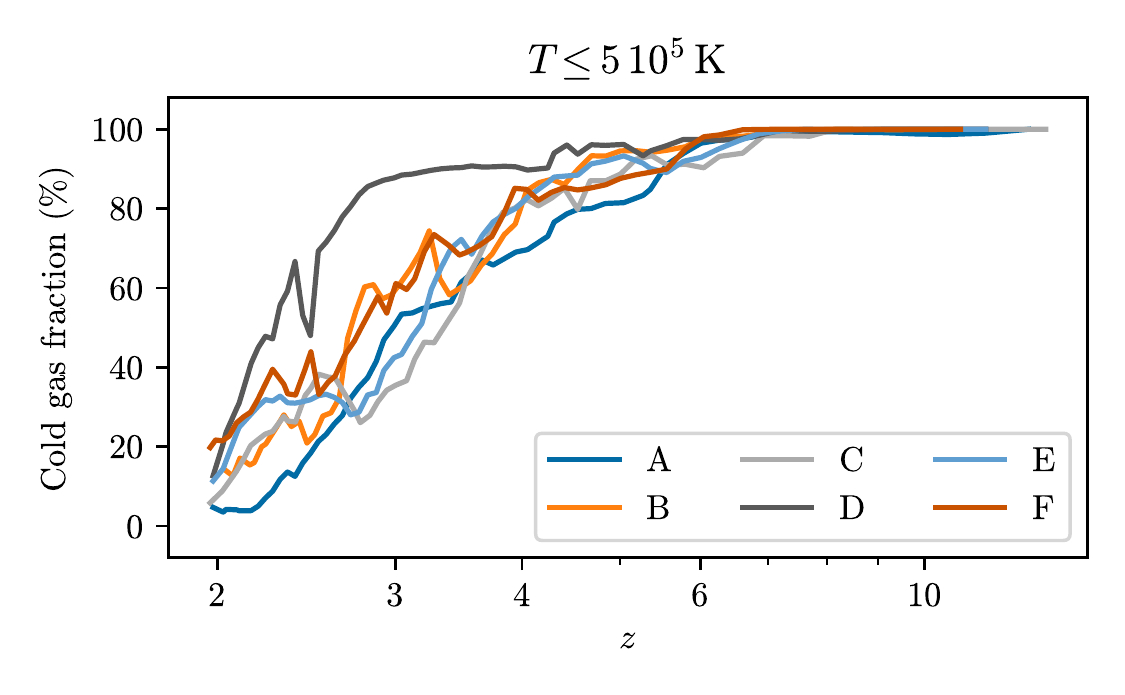}
    \includegraphics[width=\columnwidth]{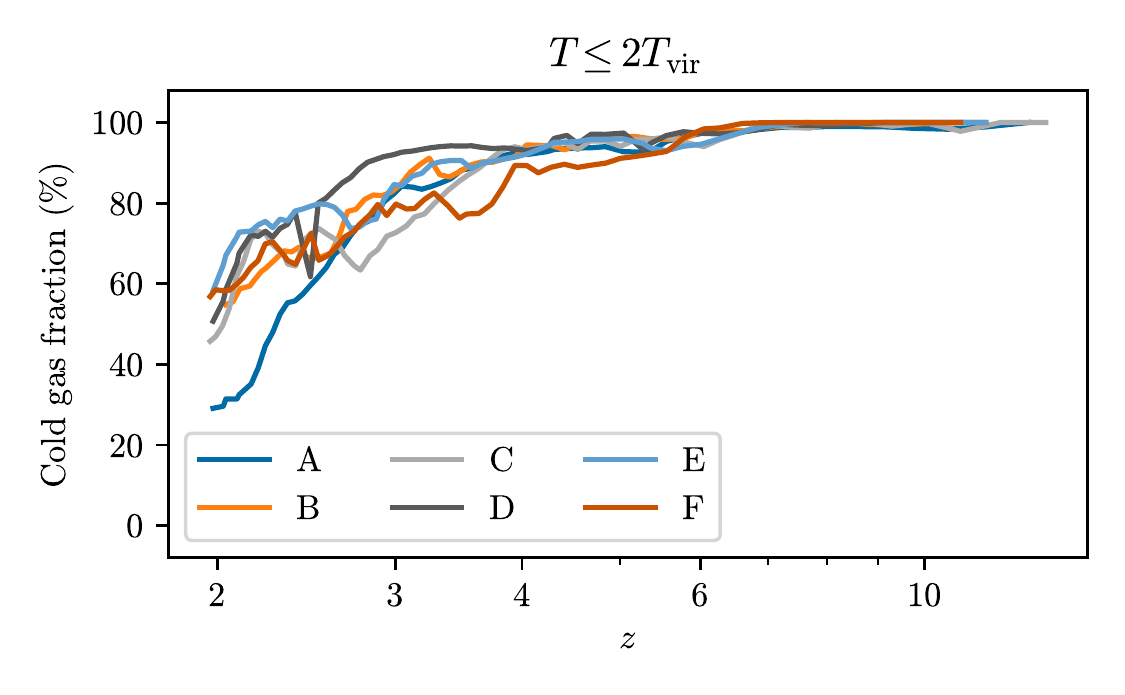}
    \caption{Evolution of the cold fraction for different temperature cut criterion.}
    \label{fig:cold_fraction_temperature_cuts}
\end{figure*}

To assess the robustness of our temperature selection, different temperature cuts have been tested.
The results are shown on \cref{fig:cold_fraction_temperature_cuts} for absolute temperature cuts (left column, from top, $T\leq \SI{e5}{K}, T\leq \SI{2.5e5}{K}$ and $T\leq \SI{5e5}{K}$) or temperature cuts relative to the virial temperature (right column, from top to bottom $T/T_\mathrm{vir}\leq 1/2, 1$ and $ 2$ ).
In the range explored, all temperature cuts lead to the same conclusion that at high redshift, all the accretion is through cold flows.
For the mass range presented in this paper, the cold gas fraction decreases to a few percents by $z=2$.
Finally, our results are qualitatively unchanged when using a different temperature cut, though the difference in evolution between cold and hot mode becomes blurred with lower thresholds.

\end{document}